\documentstyle[psfig]{mn}
\def\d{\mbox{d}}
\def\G{{\cal G}}

\def\myr{$\,M_\odot\,$yr$^{-1}$}
\def\Myr{\,M_\odot\,{\mathrm yr}^{-1}}
\def\macc{$\dot M_{\mathrm acc}$}
\def\mloss{$\dot M_{\mathrm loss}$}
\def\Mloss{\dot M_{\mathrm loss}}
\def\Macc{\dot M_{\mathrm acc}}
\def\lacc{$L_{\mathrm acc}$}
\def\Lacc{L_{\mathrm acc}}
\def\h2{${}^2$H}

\def\msun{$M_{\odot}$}
\def\Msun{M_{\odot}}
\def\rsun{$R_{\odot}$}

\def\lsun{$L_{\odot}$}

\def\chem#1{${}^{#1}$}
\def\kms{$\mathrm km\,s^{-1}$}
\def\vsini{{\it v}\,sin\,{\it i}\ }

\title{The accretion of planets and brown dwarfs by giant stars -- II.  solar
mass stars on the red giant branch}
 
\author[Lionel~Siess and Mario~Livio]{Lionel~Siess$^{1,2}$ and
Mario~Livio$^1$\\ $^1$Space Telescope Science Institute, 3700 San Martin
drive, Baltimore, MD 21218 \\ $^2$Laboratoire d'Astrophysique de
l'Observatoire de Grenoble, Universit\'e Joseph Fourier, B.P.53X, F-38041,
Grenoble Cedex, France}
  
\begin{document}

\maketitle

\begin{abstract}

This paper extends our previous study of planet/brown dwarf accretion by
giant stars to solar mass stars located on the red giant branch. The model
assumes that the planet is dissipated at the bottom of the convective
envelope of the giant star. The giant's evolution is then followed in
detail. We analyze the effects of different accretion rates and different
initial conditions. The computations indicate that the accretion process is
accompanied by a substantial expansion of the star, and in the case of high
accretion rates, hot bottom burning can be activated.  The possible
observational signatures that accompany the engulfing of a planet are also
extensively investigated. They include~: the ejection of a shell and a
subsequent phase of IR emission, an increase in the ${}^7$Li surface
abundance and a potential stellar metallicity enrichment, spin-up of the
star due to the deposition of orbital angular momentum, the possible
generation of magnetic fields and a related X-ray activity due to the
development of shear at the base of the convective envelope, and the
effects on the morphology of the horizontal branch in globular clusters. We
propose that the IR excess and high Li abundance observed in 4-8\% of the G
and K giants originate from the accretion of a giant planet, a brown dwarf
or a very low-mass star.

\end{abstract}


\section{Introduction}

Planets and brown dwarfs have now been detected around several main
sequence stars (e.g. Mayor \& Queloz 1995, Butler \& Marcy 1996, Cochran et
al.  1997, Rebolo et al. 1995, 1996, Basri et al. 1996, Marcy et al. 1998,
Butler et al. 1998, Delfosse et al. 1998), and strikingly, a fraction of
these ``hot Jupiters'' orbit very close to the central star, at less than 1
AU. This proximity to the central star poses a theoretical problem
regarding the mechanism that brought these planets to these orbits, but it
also raises the question of their fate. Indeed, Rasio et al. (1996) showed
that the orbit of 51 Peg's companion is tidally unstable and that the
planet will ultimately be engulfed in the stellar envelope. The problem we
address here is that of the effects of the accretion of a giant planet or a
brown dwarf on the structure and evolution of solar mass stars on the red
giant branch. This
study is the continuation of our previous work (Siess \& Livio 1998), in
which we investigated the swallowing of massive planets/brown dwarfs by a
3\msun\ AGB star.

The general scenario for the accretion process is as follows~: planets in
close orbits (as observed) will be engulfed in the envelope of giant stars
as the latter evolve away from the main sequence. Due to viscous and tidal
forces, angular momentum is imparted to the envelope and the planet
spirals-in. Previous studies of the spiralling-in process (e.g. Livio \&
Soker 1984; Soker, Harpaz \& Livio 1984) have revealed that depending on
the mass of the planet/brown dwarf three different evolutionary paths can
take place. If the mass of the planet is larger than some critical value
$M_{\mathrm{crit}}$, the planet can accrete a fraction of the envelope
mass. On the other hand, lower mass planets will either evaporate or
collide with the stellar core.  For the specific stellar model used in
their computation, Livio and Soker (1984) estimated $M_{\mathrm{crit}} \sim
20 M_{\mathrm{Jup}}$.  However, these results must be taken with great
caution since the physical processes were treated only approximatively. We
also emphasize that such computations are very sensitive to the stellar
model, notably to the size and mass of the convective envelope. More
recently, Sandquist et al. (1998) performed 3D hydrodynamical simulations
of Jupiter-like planets impacting onto solar type main sequence stars. They
also found that depending on the initial stellar model, planets can be
evaporated or survive the crossing of the convective zone.

The present work has been motivated by the fact that planets surrounding
low mass stars are more likely to be engulfed in the star's envelope during
the red giant branch (RGB) phase than during the asymptotic giant branch
(AGB) phase. Indeed, if we compare (for different stars) the ratio of the
maximum radius a star can reach during the RGB to its maximum radius on the
AGB phase, this ratio is much closer to one for low mass stars ($M \la 2$\msun)
than more massive stars (e.g. Soker 1998a). Consequently, we expect that in
most cases low mass stars will interact with close low-mass binary
companions on the RGB.

In the present paper, we used a spherically symmetric stellar evolution
code to follow the evolution of a solar type star that accretes a giant
planet or a brown dwarf on the red giant branch. In $\S$\ref{conditions},
we describe briefly the computational approach and the initial
models. Then, in $\S$\ref{struc}, we analyze the effects of accretion on
the structure and the evolution of the star. We present results for
different mass accretion rates and different initial stellar models on the
RGB. The observational signatures expected from this process are
investigated in detail in $\S$\ref{signature}. A discussion and conclusions
follow.

\section{Initial conditions}
\label{conditions}

To investigate the effects of planet accretion by solar type stars, we used
3 different evolutionary models of the sun corresponding to 3 different
points on the RGB. These models are characterized by their mass, which is
equal to $M = 1.0$, 0.9 and 0.8\msun\ and they are referred to as model
\#A, \#B and \#C, respectively. The main properties of the initial models
are summarized in Table \ref{tab1}. The general RGB structure is
a degenerated He core, surrounded by a thin hydrogen burning
shell (HBS) that provides most of the luminosity. A thin radiative layer
separates the HBS from the large convective envelope. The star was assumed
to lose mass as prescribed by Reimer's law (1975) at a rate given by
\begin{equation}
\Mloss = -3.98 \times 10^{-13}\, \eta_R \frac{L\,R}{M}\,\Myr\ ,
\end{equation}
where the parameter $\eta_R = 0.5$ in our computations, and $L$, $R$, and
$M$ are given in solar units. 

The input physics for the accretion process has been previously described
by Siess and Livio (1998). Briefly, to determine the locus of the
dissipation process we estimated the Virial temperature of the brown dwarf
and the critical radius at which tidal effects lead to a total disruption
of the brown dwarf. For characteristic values of the mass and radius for
the brown dwarf, we found a temperature of a few million K and a critical
radius of the order of $\sim 0.1-1$\rsun. From the numbers presented in
Table \ref{tab1}, we can localize the accretion process at the bottom of
the convective envelope near the HBS. To estimate the accretion rate \macc,
we determined the orbital decay timescale of the brown dwarf in the
vicinity of the dissipation region. This timescale roughly corresponds to
the time required by the brown dwarf to cross the dissipation region and
thus represents the rate at which mass is added there. For typical stellar
parameters, we inferred mass accretion rates of the order of $10^{-4} -
10^{-5}$\myr, depending mainly on the mass of the brown dwarf.

\begin{table}
\caption{Physical properties of the initial models}
\label{tab1}
\begin{tabular}{@{}lcccccc}
  & \#A  & \#B & \#C \\
Mass (\msun) & 0.9928 & 0.9004 & 0.8072 \\
Radius (\rsun) & 10.72 & 88.71 & 137.10 \\
L (\lsun) & 45.2 & 1133.6 & 2016.3 \\
age $t_0$ ($10^9$\,yr) & 12.1398325 & 12.1999969 & 12.2022346 \\
$k_{\mathrm env} ^\dag$ & 0.3324 & 0.2366 & 0.2098 \\ 
M$_{\mathrm env,bot}$ (\msun) & 0.2583 & 0.3904 & 0.4336 \\
M$_{\mathrm HBS,top}$ (\msun) & 0.2474 & 0.3883 & 0.4322 \\
M$_{\mathrm core}$ (\msun) & 0.2419 & 0.3867 & 0.4310 \\
R$_{\mathrm env,bot}$ (\rsun) & 0.543 & 0.791 & 0.831 \\
R$_{\mathrm HBS,top}$ (\rsun) & 0.08586 & 0.1033 & 0.1060\\
R$_{\mathrm core}$ (\rsun) & 0.02587 & 0.02538 & 0.02492 \\
$\rho_{\mathrm env,bot}$ (g\,cm$^{-3}$) & 4.057$\,10^{-2}$ &
1.586$\,10^{-3}$ & 8.715$\,10^{-4}$ \\ 
$\rho_{\mathrm HBS,top}$ (g\,cm$^{-3}$) & 2.733 & 0.3726 & 0.2510 \\
$\rho_{\mathrm core}$ (g\,cm$^{-3}$) & 369.7 & 135.2 & 112.0 \\
T$_{\mathrm env,bot}$ ($10^6$ K) & 3.286 & 2.324 & 2.209 \\
T$_{\mathrm HBS,top}$ ($10^6$ K) & 11.0 & 12.3 & 12.6 \\
T$_{\mathrm core}$ ($10^6$ K) & 29.8 & 44.5 & 48.2 \\
T$_{\mathrm c}$ ($10^6$ K)$^\ddag$ & 34.6 & 60.9 & 69.9 \\
\end{tabular}

$^\dag$gyration radius [$(I_{\mathrm env}/M_{\mathrm env}R_{\mathrm
    env}^2){1/2}$]\\
$^\ddag$central temperature\\
\end{table} 

During the accretion phase, we adopted a uniform mass deposition profile as
described in Siess and Livio (1998). More specifically, we deposit in each
numerical shell an amount of accreted matter proportional to the mass of
the shell, starting from the top of the HBS and up to a radius $r =
1$\rsun. Our simulations indicate that small changes in the profile of mass
deposition do not substantially affect the results. The main point here is
the depth at which the mass is deposited inside the star, and in the following
sections we identify the {\sl accretion zone} as the region located close
to the base of the convective envelope where most of the energy
accompanying the accretion process is released. Finally, if the convective
envelope reaches the HBS (beginning of Hot Bottom Burning or HBB), the
matter is deposited from the base of the convective envelope (which is
inside the HBS).  In our computations, we also assumed that the matter is
deposited with the same entropy as the one of the local matter,
implying that the rotational energy has been used to heat up the accreted
matter (e.g. Harpaz \& Soker 1994). The deposition of accreted matter is
also accompanied by modifications of the chemical abundances. We adopted a
solar metallicity with relative abundances given by Anders and Grevesse
(1989) for the accreted matter and we assumed that deuterium had been
previously burnt in the brown dwarf. Finally, for numerical convenience, we
turned off the mass loss during the accretion phase and used grey
atmosphere models to speed up the convergence process.

\section{Numerical results}
\label{struc}

The stellar evolutionary code used in these computations has been described
in detail in several papers (e.g. Forestini \& Charbonnel 1997, Siess et
al. 1997) and we refer the reader to those publications for more
details.  Computations are presented for 2 characteristic accretion rates
of $10^{-4}$ and $10^{-5}$\myr, and they involve different masses for the
planet/brown dwarf (see Table \ref{model}).

\begin{table}
\caption{Numerical models}
\label{model}
\begin{tabular}{@{}lccc}
Models & $M_{\mathrm ini}$ ($M_\odot$) & \macc (\myr) & accreted mass
($M_\odot$) \\
A0 & 0.99 & $10^{-5}$ & 0.10 \\
A1 & 0.99 & $10^{-5}$ & 0.01 \\
A2 & 0.99 & $10^{-4}$ & 0.10 \\
A3 & 0.99 & $10^{-4}$ & 0.01 \\
B0 & 0.90 & $10^{-5}$ & 0.10 \\
B1 & 0.90 & $10^{-5}$ & 0.01 \\
B2 & 0.90 & $10^{-4}$ & 0.0051 \\
B3$^\dag$ & 0.90 & $10^{-4}$ & 0.0058 \\
C0 & 0.81 & $10^{-5}$ & 0.10 \\
C1 & 0.81 & $10^{-5}$ & 0.01 \\
C2 & 0.81 & $10^{-4}$ & 0.0010 \\
C3 & 0.81 & $10^{-4}$ & 0.0025 \\
C4$^\dag$ & 0.81 & $10^{-4}$ & 0.0030 \\
\end{tabular}

$^\dag$For these models, we were not able to follow reliably the
evolution and the computations were stopped at the indicated accreted mass 
\end{table} 

\subsection{Evolution of the structure for lower accretion rates}

\begin{figure}
\psfig{file=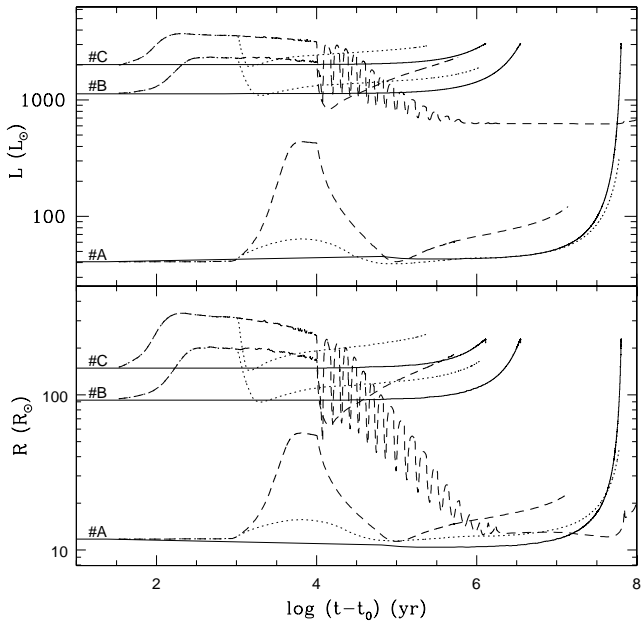,width=\columnwidth} %,height=\columnwidth}
\caption[]{The evolution of the radius $R$ and luminosity $L$ for the
different models computed with a lower accretion rate ($\Macc =
10^{-5}$\myr). The time $t_0$ corresponds to the age of the
initial model and is indicated for the different models in Table
\ref{tab1}. The solid lines refer to the standard solar model (no
accretion), the dotted and dashed lines correspond to models in which the
accreted mass is equal to 0.01 and 0.1\msun, respectively. The depicted
models are A0, A1, B0, B1, C0 and C1.}
\label{evol5}
\end{figure}

The deposition of mass in the star is accompanied by the release of a large
amount of gravitational energy that acts to expand the envelope
(Fig. \ref{evol5}).  However, because of the long timescale of heat
diffusion, the expansion is delayed, especially for the low luminosity
stars (cases A0 and A1). As the radius increases, the star also becomes
more luminous. According to Reimer's law, these modifications to the
structure significantly increase the mass loss rate, and we expect some
changes in the subsequent evolution and stellar environment (see
$\S$\ref{shell} and $\S$\ref{morpho}). The expansion proceeds until a
thermal equilibrium is reached, in which the rate of energy loss by the
surface balances the rate of energy deposition by the accretion
process. Depending on the initial model, the return to contraction occurs
after $\sim 6200$, $\sim 470$ and $\sim 200$\,yr, in cases A0, B0 and C0,
respectively. This timescale corresponds roughly to the time of heat
diffusion (thermal timescale) of the expanding part of the star and it is
approximatively given by $\tau_{\mathrm therm} \sim \G M M_{\mathrm env}/
2 R \Lacc$, where $M_{\mathrm env}$ is the mass of the envelope and \lacc\ 
is the accretion luminosity. The latter is given by $\Lacc(r)=\G M_r
\Macc/r$, where \macc\ is the mass accretion rate and $M_r$ and $r$ the
mass and radial coordinates of the accretion zone.
It should be noted that \lacc\ represents an upper limit to the accretion
luminosity since energy is lost during the spiralling-in process to heating
and spinning up the envelope.  However, throughout most of the envelope,
the orbital energy disspation rate represents a small fraction of the
intrinsic stellar luminosity and thus we do not expect that the stellar
structure will be significantly affected until the planet reaches the
accretion zone. Finally, we should mention that the energy deposition might
not be local (in the accretion region). For example, the star could be
tidally excited, oscillations may be initiated and then damped elsewhere in
the envelope. Also, in this common envelope phase, magnetic fields might be
generated in the accretion region (see $\S$\ref{xray}) and energy could be
dissipated at the surface of the star or in the corona (e.g.  Regos \& Tout
1995).

\begin{figure}
\psfig{file=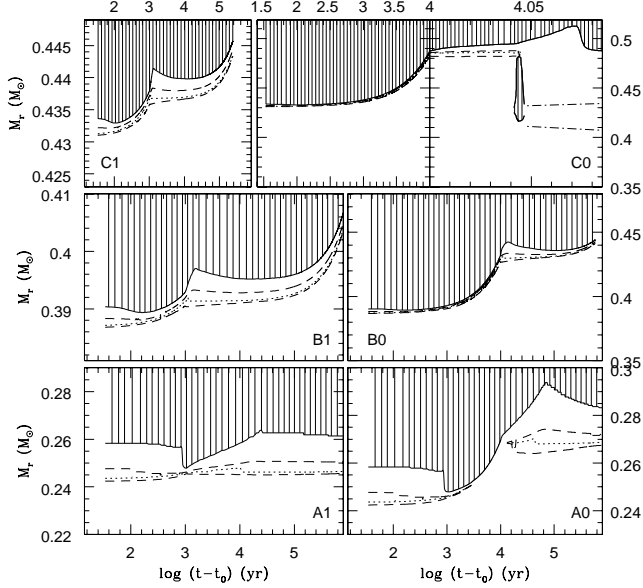,width=\columnwidth}
\caption[]{The evolution of the boundaries of the HBS and the convective
envelope. The dashed lines delineate the burning shells
($\varepsilon_{\mathrm nuc} > 5$ erg\,g${}^{-1}$s${}^{-1}$) and the dotted
line indicates the locus of maximum energy generation. The hatched areas
indicate convective regions. The different models are labeled in each panel
and were computed with an accretion rate equal to $10^{-5}$\myr. Note the
extinction of the HBS in case A0 (lower right panel). In the upper right
panel (case C0), a thermal instability develops and leads to the
development of a convective tongue. The dotted-dashed lines delineate the
He burning shell.}
\label{kippen}
\end{figure}

During the envelope expansion, the opacity $\kappa$ increases in the
accretion region and rapidly the radiative gradient becomes larger than the
adiabatic one ($\nabla_{\mathrm rad} \propto \kappa L >
\nabla_{\mathrm{ad}}$); thus the convective envelope deepens (Fig.
\ref{kippen}).  Thereafter, as the HBS advances in mass and reaches the
cool regions that have been previously expanded, the nuclear energy
production rate decreases significantly. As an order of magnitude, during
most of the accretion phase, the luminosity due to H burning represents
only a few percent of the total stellar luminosity. The accretion
luminosity thus supports the convective envelope almost entirely and the
accretion region acts as a source shell that prevents an efficient H
burning.

\begin{figure}
\psfig{file=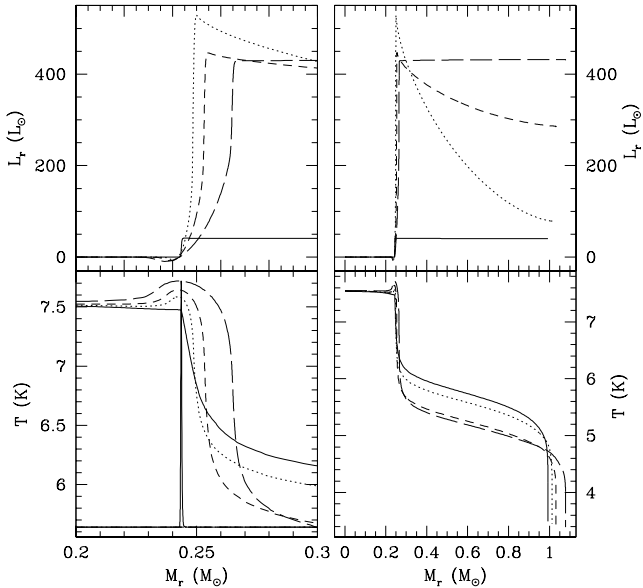,width=\columnwidth}
\caption[]{The evolution of the profiles of luminosity ($L_r$) and
temperature ($T$) for case A0. The solid line refers to the initial model,
the dotted, short-dashed and long-dashed lines correspond to models for
which 0.019, 0.039 and 0.086\msun\ have been accreted, respectively. The
accretion rate is equal to $10^{-5}$\myr. The peak in the lower left panel
represents the nuclear energy production rate in the initial model and thus
delineates the H burning shell.}
\label{prof1}
\end{figure}

To see how the structure is affected by the accretion process, we have
presented in Fig. \ref{prof1} the profiles of several variables for case
A0, but they are globally similar in the other models. In the accretion
region where a large amount of energy is deposited, the luminosity and
temperature rise. But, due to the relatively long timescale associated with
heat diffusion, energy accumulates locally and both profiles now present a
bump. The heat excess finally diffuses when the star has sufficiently
expanded and a monotonically increasing luminosity profile is
restored. However, the temperature profile still presents a local maximum
that persists throughout the accretion phase. This is due to the fact that
this region undergoes a compressional heating from the mass accumulated
above it and also from the expanding region surrounding the core. Indeed,
the luminosity $L_r$ is negative just behind the luminosity jump,
indicating a region in expansion ($\varepsilon_{\mathrm grav} < 0$). The
temperature is thus compelled to rise and, as we shall see in
$\S$\ref{relax}, this situation can give rise to a thermal instability.
The temperature gradient also becomes progressively steeper beyond the bump
(Fig. \ref{prof1}) as energy accumulates. This contributes to the shrinkage
of the HBS and to the subsequent drop in the nuclear luminosity.

\begin{figure}
\psfig{file=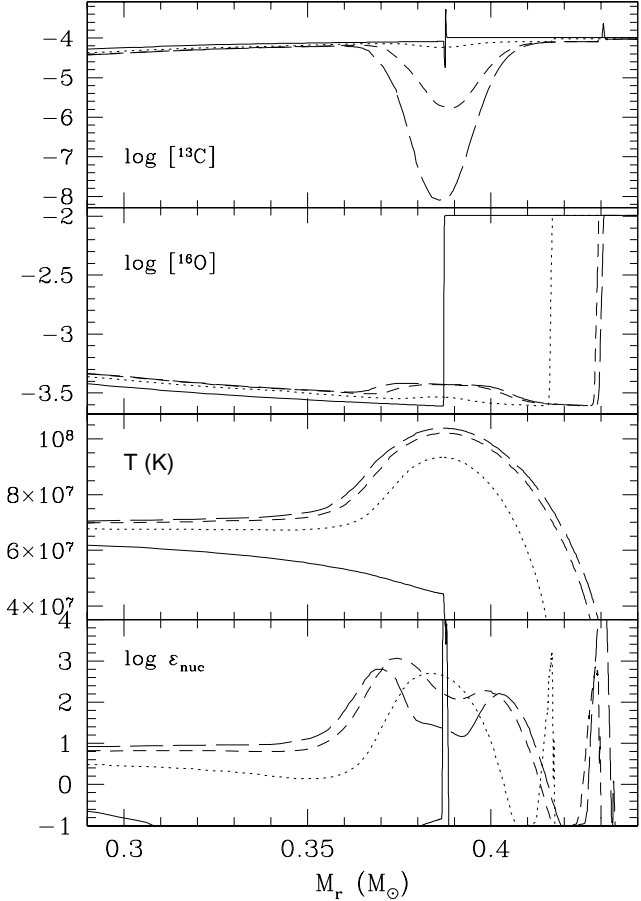,width=\columnwidth}
\caption[]{Evolution of the nuclear energy production rate
($\varepsilon_{\mathrm nuc}$), temperature (T), and \chem{16}O and
\chem{13}C chemical abundances for model B0 ($\Macc = 10^{-5}$\myr). The
solid lines refer to the initial model $t = t_0$. The dotted, short-dashed
and long-dashed lines correspond to an age $t-t_0$ equal to 6791, 9760, and
11040 yr, respectively. Note the persistence of the temperature excess
during the accretion phase.}
\label{prof2}
\end{figure}

In the HBS, prior to accretion, the nuclear energy production rate is
mainly supplied by the CNO bi-cycle. More precisely, the most important
reactions (and their contribution) are \chem{13}C(p,$\gamma$)\chem{14}N
($\sim 25\%$), \chem{14}N(p,$\gamma$)\chem{15}O ($\sim 23\%
$),\chem{15}N(p,$\alpha $)\chem{12}C ($\sim 16\%$) and
\chem{15}O($\beta^+,\nu$)\chem{15}N ($\sim 16\%$). However, because the HBS
advances in mass into a cooler region, the pp1 branch progressively takes
over the nuclear energy production and during the accretion phase, the
reactions 2\,\chem{3}He($\gamma $,2p)$\alpha$ and
\chem{13}C(p,$\gamma$)\chem{14}N contribute $\sim 45$\% and $\sim 10$\% of
the total nuclear energy production, respectively. On the other hand, in
the region of maximum temperature, \chem{13}C is now efficiently burnt
through the \chem{13}C($\alpha $,n)\chem{16}O reaction. This reaction is
highly energetic and provides up to $\sim 20\%$ of the global nuclear energy
production (which still remains very low, however). In Fig. \ref{prof2}, we
show the evolution of the nuclear activity for case B0. First, \chem{13}C
burning is ignited in the vicinity of the temperature peak (dotted
line). Then, as it is depleted and converted into \chem{16}O, the nuclear
energy production rate progressively decreases near the temperature peak
(short-dashed and long-dashed curves), and finally we end up with a
configuration in which \chem{13}C is burnt through
\chem{13}C($\alpha$,n)\chem{16}O around the temperature peak and through
\chem{13}C(p,$\gamma$)\chem{14}N above the bump in the HBS. In case C0, the
temperature in the peak reaches $\sim 1.2\,10^8$K and He is ignited
off-center in a non degenerate shell (we will discuss this special case
further in $\S$\ref{relax}).

\subsection{Higher accretion rates}

\begin{figure}
\psfig{file=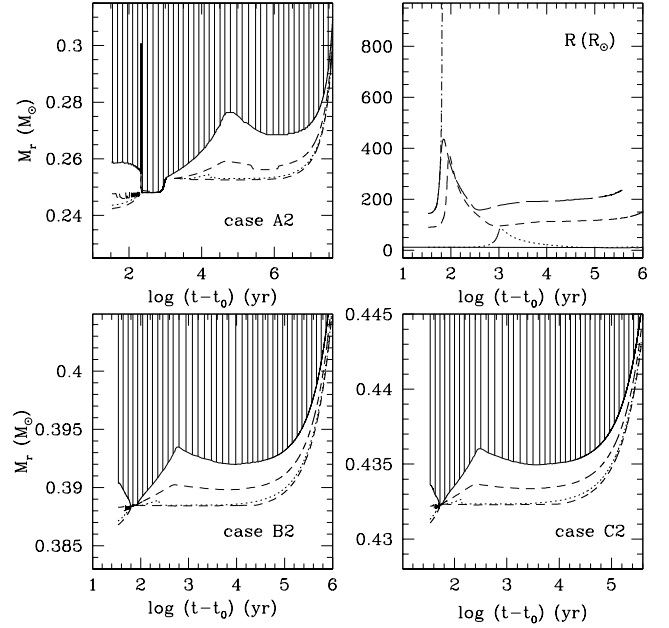,width=\columnwidth} %,height=\columnwidth}
\caption[]{The evolution of the internal structure (left and lower right
panels) and radius (upper right panel) for the high accretion rate case
($\Macc = 10^{-4}\Myr$). The evolution of the structure is depicted in a
similar way to that in Fig. \ref{kippen}. Note the appearance of hot bottom
burning in these computations. In the upper right panel, the solid, dotted,
short-, long- and dot-dashed lines refer to the standard case and to cases
A2, B2, C2 and C4, respectively.}
\label{struc4}
\end{figure}

For higher accretion rates, the situation is somewhat different because of
the appearance of hot bottom burning (HBB). Indeed, contrary to the
lower accretion rate case, the larger energy release contributes
to a further increase in the opacity and the convective envelope deepens
even more. It finally reaches the HBS where its advance stops
(Fig. \ref{struc4}). Our computations indicate that HBB is activated
approximately 200, 25 and 15 yr after the onset of accretion in cases A2,
B2 and C2 respectively. The temperature at the base of the convective
envelope rises suddenly from $\sim 2-3$ to about $13-18$ million Kelvin,
depending on the initial model. However, this temperature is still too low
to give rise to a rich nucleosynthesis.
The onset of HBB entails a significant and fast expansion of the envelope
as heat is directly transferred to the convective zone. Moreover, the
arrival of \chem{3}He from the convective envelope and its destruction
through the reaction 2\,\chem{3}He($\gamma$,2p)$\alpha$ increases
temporarily the nuclear luminosity in the HBS. During the HBB phase,
\chem{7}Li is slowly depleted because of the competing effects of Li
production through the reactions
$\alpha$(\chem{3}He,$\gamma$)\chem{7}Be($e^-,\nu$)\chem{7}Li and Li
destruction through the reaction \chem{7}Li(p,$\alpha$)$\alpha$ (see also
$\S$\ref{chem}).

Finally, in cases B2 and C2 we were not able to follow reliably the
computations.  The star's tendency to expand and increase its luminosity
made the computations intractable. At that point the envelope is probably
ejected and the star enters a proto-planetary nebula phase.

\subsection{The relaxation of the star}
\label{relax}

After the end of the accretion phase, the envelope rapidly contracts
because of the sudden suppression of accretion energy. However, in the
models in which the duration of the accretion process is shorter than the
thermal timescale associated with the envelope (as in cases B1 and C1 for
example), the expansion still proceeds for a while after the end of mass
addition. Thereafter, the gravitational energy released during the
contraction phase heats up the central parts of the star and the HBS
re-ignites. During that period of contraction, the convective envelope
retreats towards the surface and HBB disappears. The contraction rate is
very fast, especially for stars having a shallower convective
envelope. The computations indicate that in a few thousand years, the radius
decreases by a factor of $\sim 2-3$ which, we suspect, could initiate
stellar pulsations. When the HBS finally provides most of the stellar
luminosity ($\ga 80$\%), the contraction ends and the star resumes a standard
evolution, climbing the red giant branch until He flashes at its tip.
Interestingly, the star is now more luminous and has a larger radius than
in the standard evolution despite the small amount of added mass 
(Fig. \ref{evol5}). The enhanced mass loss rate that follows can affect the
subsequent evolution (see $\S$\ref{morpho}).

\begin{figure}
\psfig{file=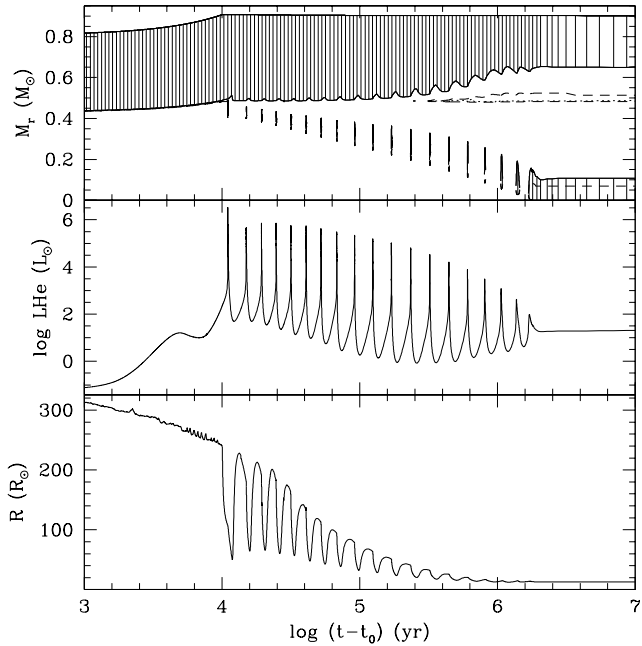,width=\columnwidth} %,height=\columnwidth}
\caption[]{The evolution of the radius, He luminosity and internal
structure for case C0. In the upper panel, the different lines follow the
same convention as in Fig. \ref{kippen}.}
\label{pulse}
\end{figure}

In all our simulations, except for case C0, the temperature ``bump'' 
dissipated as the gravitational energy released during the contraction
phase heated up the stellar interior. However in this particular model, in
which the accretion process is maintained for $\sim 10^4$yr at a rate equal
to $10^{-5}$\myr, the temperature in the peak reaches $\sim 1.2\,10^8$\,K and
He ignites off-center in a non degenerate shell. The ignition of He
through the $3\alpha$ reactions is very energetic and this gives rise to a
thermal instability (Fig. \ref{pulse}).
The geometrical thinness and the large opacity of the He burning shell
(HeBS) are responsible for the thermal runaway that leads to the
development of a convective tongue. The release of energy due to He burning
is used to expand the layers above and secondarily to heat up the core. The
HBS is pushed outward into a cooler region and it extinguishes. The
remaining energy excess is then absorbed by the envelope which subsequently
expands. During the thermal pulse, the nuclear energy production comes from
3$\alpha$ ($\sim 70$\%), \chem{14}N($\alpha, \gamma$)\chem{18}F ($\sim
17\%$) and, to a lesser extent, from \chem{18}O($\alpha,\gamma$)\chem{22}Ne
and \chem{18}O($\alpha,\gamma $)\chem{22}N ($\sim 7\%$ each). The thermal
pulse is also responsible for the further destruction of H through
\chem{12}C(p,$\gamma$)\chem{13}N.  After a significant expansion that
allows the star to get rid of the energy excess, contraction
resumes. However, due to the extremely low H content left after the pulse
(the H mass fraction $X <10^{-20}$), the HBS cannot ignite before a new
thermal instability is triggered. A new He shell flash thus occurs deeper
inside the star where both the temperature and He abundance are higher.  As
evolution proceeds, the pulse duration increases but the maximum luminosity
due to He burning decreases mainly because of the progressive lifting of
the core degeneracy as the temperature increases. During the subsequent
evolution, the star experiences relatively short phases of expansion
following the thermal pulses and longer phases of contraction. The
deepening of the pulse accounts for the smaller and smaller increase in
radius following the He shell flashes since most of the released energy is
absorbed in the radiative layers before reaching the convective envelope.
After the 12$^{th}$ pulse, the HBS ignites precisely at the mass coordinate
corresponding to the top of the first convective tongue in which the H
abundance is high ($X \gg 10^{-20}$). The H luminosity then increases and
the thermal pulse finally reaches the center of the star. Because of the
previous thermal instabilities, the degeneracy has been lifted in the
interior and consequently He does not flash. Helium is now burnt at the
center and the star will spend the next $\sim 4\,10^7$yr on the Horizontal
branch.

\section{Potential observational signatures of an accretion event}
\label{signature}

In this section, we attempt to give a broad overview of the different
observational signatures that might be expected from the accretion of a
planet or a brown dwarf by a giant star on the red giant branch. In
general, effects may be expected in at least six different areas, which we
describe in some detail below. The modifications to the structure during
the accretion process increase the mass loss rate and may lead to the
ejection of shells. They also have potential effects on the morphology of
the horizontal branch. The deposition of angular momentum during the
spiralling-in process can also alter the stellar structure by increasing
the rotation rate and this may also produce X-ray emission. Finally, the
mixing of the dissipated planet material in the envelope can change the
surface chemical abundances.  All of these effects may have important
observational consequences but, as we shall see in the following sections,
it is more the combination of several signatures, such as a high rotation
rate and/or a large IR excess and/or a high lithium abundance that can
identify an accretion event less ambiguously,  rather than the detection of
one particular signature.

\subsection{Modifications of the photospheric chemical composition}
\label{chem}

During the red giant phase, chemical changes are mainly the result of
``dilution'' effects due to the deepening of the convective envelope but
light elements such as \chem{7}Li, \chem{9}Be and \chem{11}B can also be
burnt as the convective envelope approaches its deepest extent. The first
dredge up reduces the \chem{7}Li abundance by more than 2 orders of
magnitude. There is also a decrease in the \chem{12}C/\chem{13}C ratio by a
factor of 3, changing from about 83 to 26. The first dredge up also
accounts for the large enrichment of \chem{3}He as a by product of H
burning.

\begin{figure}
\psfig{file=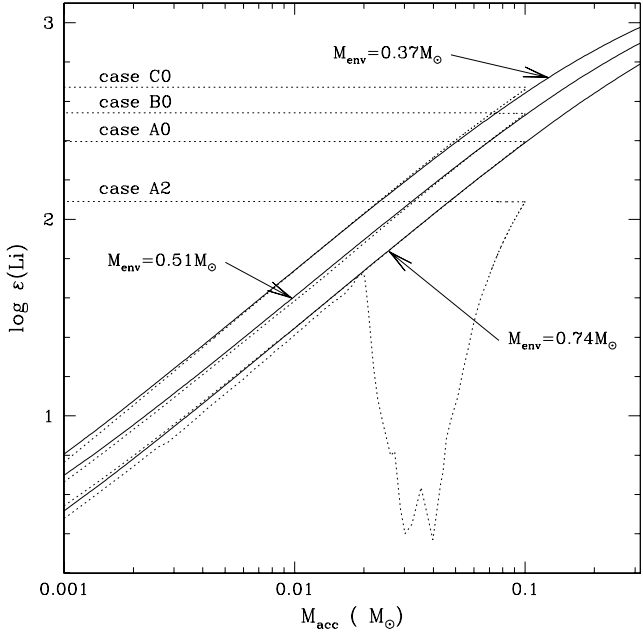,width=\columnwidth}
\caption[]{The evolution of the \chem{7}Li surface abundance as a function
of the accreted mass. The solid lines represent the \chem{7}Li abundance as
a result of dilution for the 3 initial envelope masses characterizing our
initial models ($M_{\mathrm env} = 0.74$, 0.51 and 0.37\msun\ in models A,
B and C, respectively). The dotted lines represent the results of our
computations for cases A0, B0, C0 and A2. Note the large Li depletion in
case A2 due to the long lasting HBB.}
\label{dilu}
\end{figure}

With our assumed chemical composition for giant planets and brown dwarfs
(from Anders \& Grevesse 1989), the deposition of accreted matter, mostly
in the convective envelope, will principally modify the chemical abundances
of the light elements that were previously depleted during the first dredge
up. In that respect, \chem{7}Li will be the most affected. Our computations
indicate that, depending on the accreted mass, the \chem{7}Li abundance can
be increased by more than 2 orders of magnitude, making this element easily
detectable at the surface of the star\footnote{Following de la Reza et
al. (1996), we define as a ``Li rich G or K giant'' a star with a Li
abundance larger than log$\, \varepsilon(\mathrm{Li}) = 1.2$ [where by
definition log\,$\varepsilon($H$) = 12.00$]} (see also $\S$\ref{lirich}). As
an illustration, we present in Fig. \ref{dilu} the \chem{7}Li surface
abundances 
resulting from our computations and compare them with what is expected from
simple dilution effects. In the dilution hypothesis, we assume that the
envelope mass $M_{\mathrm env}$ is held constant during the accretion
process and that no nuclear burning is taking place in the convective zone.
The mass fraction $X_i$ of an element $i$ is then simply given by
\begin{equation}
X_i = \frac{X^{\mathrm env}_i \times M_{\mathrm env}+ X^{\mathrm acc}_i
\times M_{\mathrm acc}}{M_{\mathrm acc}+M_{\mathrm env}}\ ,
\label{eqli}
\end{equation}
where $M_{\mathrm acc}$ is the mass accreted into the envelope, and
$X^{\mathrm env}_i$ and $X^{\mathrm acc}_i$ are the mass fraction of
element $i$ in the envelope and in the accreted matter, respectively. Note
that this expression is general in the sense that $X_i$ can either
represent the H mass fraction $X$, the He mass fraction $Y$ or the
metallicity $Z$. Figure \ref{dilu} shows that the simulations follow
relatively well the theoretical curves defined by Eq. (\ref{eqli}). The
slight deviations from this relation are mainly due to nuclear burning that
causes Li depletion. Note that in case A2, \chem{7}Li is severely depleted
due to the long lasting HBB.
The deposition of accreted matter will also increase the \chem{9}Be and
\chem{11}B chemical abundances, but to a lower extent, since the effects of
dilution during the first dredge up are smaller for these elements. A
modest increase in the \chem{12}C/\chem{13}C ratio can also take place.

The effects of planet accretion on the stellar metallicity are very
sensitive to the amount and chemical composition of the accreted material
and also to the mass of the convective envelope (which depends on the
evolutionary status of the star). If the metal content of the giant planet
is higher than the one adopted here, the metallicity can be further
enhanced. Secondly, if the planet is engulfed while the star is still on
the main sequence for example, the convective envelope is then very shallow
($M_{\mathrm env} \simeq 0.02$\msun\ for the sun), and consequently the
stellar photospheric composition can be substantially modified
(Eq. \ref{eqli}). Recent determinations of the chemical composition of the
main sequence stars hosting planets (Gonzales 1997, 1998) indeed show that
in some cases the metallicity is abnormally high, suggesting that
proto-planetary material or planets have been accreted (e.g. Lin et
al. 1996, Laughlin \& Adams 1997). More recently, Sandquist et al. (1998)
showed that a Jupiter-like planet can be evaporated in the convective
envelope of the sun. They also inferred that in some cases a metal
enrichment as high as 50\% can be obtained, but the exact value depends
greatly on the mass of the convective zone and on the chemical composition
and interior structure of the giant planet.

All the above suggests that if the presence of planets around solar type
stars is rather common, and if some of these planets are indeed in close
orbits, then certain chemical peculiarities in giants stars may be the
consequence of the accretion of planets.

\subsection{Rotation}
\label{rotation}

During the spiralling-in process, angular momentum is imparted to the
envelope. As a consequence, the stellar rotation rate will be substantially
increased. More quantitatively, if all the orbital angular momentum of the
planet is entirely deposited and redistributed in the envelope, i.e. if we
assume that the convective envelope is rotating rigidly, then the ratio of
the envelope angular velocity $\omega$ to the critical velocity (surface
Keplerian angular velocity) $\omega_K$ is given by
\begin{equation}
\frac{\omega}{\omega_K} \approx 0.10 \Bigl(\frac{M_{\mathrm bd}}{0.01
\Msun}\Bigr) \Bigl(\frac{M_{\mathrm{env}}}{\Msun} \Bigr)^{-1}
\Bigl(\frac{k^2}{0.1} \Bigr)^{-1} \Bigl(\frac{a}{R}\Bigr)^{1/2}\ , 
\label{omega}
\end{equation}
where $a$ is the orbital separation of the brown dwarf, $k$ the gyration
radius and $M_{\mathrm{env}}$ the mass of the envelope. Equation
(\ref{omega}) indicates that only massive planets or brown dwarfs
($M_{\mathrm{bd}} > 5 M_{\mathrm{Jup}}$) can spin-up the giant envelope to
a significant fraction of the Keplerian velocity ($\ga 10$\%). The spin-up
would be most effective at the tip of the giant branch, where both
$M_{\mathrm env}$ and $k$ are smaller.

Measurements of the projected rotational velocity of luminous class III
giants indicate a sudden drop in \vsini near spectral type G0 III (Gray
1989). The same discontinuity is also observed in subgiants at spectral
type F8 IV (Gray \& Nagar 1985) and on the main sequence near mid-F stars
(Kraft 1970). These observations thus delineate in the HR diagram a
rotational dividing line (RDL) where slow rotators ($v\,\sin\,i \la 5$
\kms) populate the cool side (de Medeiros et al. 1996b). The strong braking
at the rotational discontinuity occurs very rapidly, perhaps on a time
scale as short as $\sim 10^5$ yr or less (Ayres et al. 1998), and probably
results from an efficient magnetic braking attributed to a dynamo action
(Gray 1981, 1982). In the ``rotostat'' hypothesis (Gray 1986), angular
momentum is dredged up from the rapidly rotating core by the deepening of
the convective envelope. This activates the dynamo and then an efficient
magnetic braking takes place. In this model, the rotation rate is
controlled in a self-regulatory way by the balance between the arrival of
angular momentum from the core and the strong braking that ensues, and
consequently, the rotational velocity is maintained constant. However, a
few stars do not conform to this scheme and still rotate rapidly beyond the
RDL, challenging the rotostat hypothesis. For example the case of 7 Boo =
HR 5225 remains unexplained (Gray 1989, 1991). This single star of spectral
type G5III has a moderate lithium abundance log\,$\varepsilon$(Li)=1.2 and
its rotational velocity (\vsini = 14.1 \kms) is three times higher than the
average. This isolated case could fit into a scenario in which the giant
star has recently accreted a planet and its surface layers have not yet
spun down. The moderate lithium abundance would suggest the accretion of a
relatively small planet. More generally, de Medeiros et al. (1996a) found
that among a sample of $\sim 900$ stars, less than $\sim 5$\% of the late G
and K giants located to the right of the RDL have $v\,\sin\,i>5$ \kms.

On the other side (left) of the RDL, FK Comae-type stars show extreme
rotation velocities, ($\omega/\omega_K \ga 0.10$).  The general explanation
for their fast rotation assumes that these single G or K giants are
undergoing (or have just completed) coalescence with a close binary
companion (e.g. Webbink 1977, Bopp \& Stencel 1981, Livio \& Soker 1988,
Welty and Ramsey 1994). Other scenarios have also been proposed and they
generally invoke tidal synchronization in a very close binary, normal
evolution from a rapidly rotating upper main sequence star, or the dredge
up of angular momentum (e.g. Simon \& Drake 1989). But, as pointed out by
Rucinski (1990), FK Coma has a few times more ($> 8$) angular momentum than
a typical rapidly rotating main sequence star.  Therefore, some angular
momentum must have been added (accreted) to the star. Generally FK
Comae-like stars lack photospheric lithium detection. However in some
cases, e.g. HD 36705 (Rucinski 1985), HD 33798 (Fekel \& Marschall 1991) or
1E 1751+7046 (Ambruster et al. 1997), a high lithium abundance has been
observed (see also $\S$\ref{lirich}). These observations clearly support
our accretion scenario since they combine two different outcomes of the
accretion process~: a high rotation rate and a high Li abundance. However,
different scenarii have also been proposed and among them the possibility
that these stars were initially fast rotating A or F type stars that have
rapidly evolved in the Hertzsprung gap. The suggestion is then that they
recently became giants and are now depleting lithium due to the deepening
of the convective envelope (Fekel 1988). It has also been suggested (Fekel
\& Marschall 1991) that angular momentum and CNO processed material can be
dredged up from a rapidly rotating core as the convective zone deepens.  In
this scenario Li can be produced and burnt through the chain
\chem{3}He($\alpha, \gamma$)\chem{7}Be($e, \nu$)\chem{7}Li($p,
\alpha$)\chem{4}He and, if on its way to the surface Li is not entirely
depleted, then we may account for temporarily high rotation together with
enhanced lithium abundance. However, this hypothesis would imply that
essentially every star arriving to the RGB should go through a phase of
fast rotation and high Li abundance. The paucity of such candidates poses a
severe problem to this scenario and can be regarded as additional evidence
against the rotostat hypothesis.

To conclude this section, we feel that observations support the coalescence
hypothesis for the explanation of FK Comae-like stars and we stress that
rapidly rotating stars may also have accreted angular momentum from the
swallowing of a giant planet or a brown dwarf (see also Vitenbroek et
al. 1998). The planet accretion scenario can also account for enhanced Li
abundances and in that respect, HD 33798 and HD 36705 are good candidates
for such an event. Finally, we would like to point out that fast rotation
velocities, of the order of 15-40 \kms, have also been observed in
horizontal branch stars (e.g. Peterson et al. 1995, Cohen \& McCarthy
1997). Notably, Peterson et al. (1983) already mentioned that the
deposition, during the red giant phase, of the orbital angular momentum of
a planet of a few Jupiter masses could account for the fast rotational
velocities observed in horizontal branch stars. Some AGB stars also exhibit
high rotation rates (e.g. V Hydrae). For these exceptional stars, the rapid
rotation rate is thought to be due to spinning up by a companion that has
been swallowed or is in a common envelope phase (Barnbaum et
al. 1995). More generally, the deposition of angular momentum in the
envelope of evolved stars by a companion (even substellar) is
frequently invoked to explain the axisymmetrical morphology of bipolar
nebulae (e.g. Iben \& Livio 1993, Livio 1997).

\subsection{X-Ray emission}
\label{xray}

If the accretion of angular momentum can cause the star to rotate
differentially then differential rotation and convection can generate a
magnetic dynamo. The magnetic field can then confine the plasma in the
stellar chromosphere and produce X-ray emission activity. To see if this
scenario is plausible, let us first examine the observational situation.

Observations indicate the existence of an X-ray dividing line in the HR
diagram (XDL also called coronal dividing line) which delimits giant stars
with (to the left) and without coronal emission characteristic of high
temperature gas ($T> 10^6$\,K). In the HR diagram, this line is
approximatively vertical at spectral type $\sim$K3 and it spans the
luminosity classes II, III and IV (Haisch et al. 1991, 1992). Note that
bright giants of luminosity class II and Ib turn out to be X-ray sources
if exposed sufficiently deeply (Reimers et al. 1996). On the right side of
the XDL, the vast majority of the stars exhibit a very low X-ray emission
(at least below the actual detection threshold) whereas a large spread of
X-ray luminosities is observed for G and K giants of earlier spectral type
than K3 (Strassmeier et al. 1994, H\"unsch et al. 1996). The X-ray emission
originates from the stellar corona and requires a strong magnetic field in
order to confine the hot plasma, since the thermal velocity of the
particles exceeds the escape velocity. Changes in the magnetic field
configuration (due to changes in the dynamo) are generally invoked to
account for the sudden drop in X-ray emission at the XDL (e.g. Rosner et
al. 1991, 1995) but the precise reasons why the dynamo is modified remain
unclear. Recently H\"unsch and Schr\"oder (1996) presented a ``revised''
XDL that seems to follow the evolutionary track of a 1.2\msun\ star. If
this is confirmed, then evolutionary effects may account for this
demarcation line. Low mass stars, on the right of the XDL, are generally
slow rotators on the main sequence and consequently they sustain a lower
magnetic activity. Conversely, more massive stars ($M \ga 1.6$\msun) have
faster rotation rates on the main sequence and therefore they can develop
an efficient dynamo and produce a higher X-ray emission. More recently,
H\"unsch et al. (1998) looked for X-ray emission in the ``forbidden''
region beyond the XDL and they detected 4 intrinsically X-ray bright M-type
giants~: 15 Tri, HR 5512, 42 Her and HR 7547. No evidence for binarity
exists yet for 15 Tri and HR 5512 but even if the presence of a companion
is confirmed, it cannot easily explain the X-ray emission which remains
quite unusual\footnote{A compact companion such as a white dwarf can be a
relatively strong X-ray source if it accretes matter from the giant through
an accretion disk. But such a companion would have been detected in their
observations.}. In a ROSAT survey of a complete, volume limited sample of
late type stars, H\"unsch et al. (1996) also pointed out the unusual
behavior of $\delta$ Sgr = HR 6859, an apparently single, slowly rotating
but strong X-ray emitting K3 III giant. Another puzzling case is that of HR
1362 (Strassmeier et al. 1990). Despite its low rotation rate this single
G8 IV giant presents an X-ray emission that is more than one order of
magnitude higher than that expected from the empirical rotation versus
activity trend. The strong emission of HR 1362 led Stepien (1993) to
suggest that it might represent an evolved Ap star that has preserved its
strong magnetic field. On the other side (left) of the XDL, several
unusually bright X-ray sources have also been detected. They are often
referred to as $\beta$ Ceti-like stars, based on the prototype $\beta$ Ceti
(H\"unsch et al.  1996, Maggio et al. 1998), which exhibits an extremely
high X-ray luminosity and a slow rotation rate ($v\sin i \simeq 3$
\kms). These observations are puzzling since it is difficult to reconcile
the high X-ray emission with the slow rotation rate. Below we investigate
the physical effects related to the accretion of a planet/brown dwarf in
terms of the generation of magnetic field and X-ray emission.

Suppose that an initially slowly rotating star undergoes an accretion
event. As the planet spirals in, it will deposit angular momentum in the
envelope and finally dissipate close to the core. However, the resultant
profile of angular velocity $\omega(r)$ may not be uniform throughout the
envelope but rather may be steeper close to the dissipation region, where
the gravitational potential increases abruptly. To quantify this, at least
approximatively, we follow Livio and Soker (1988) and estimate the
parameter $\gamma_{\mathrm{CE}}$ defined by
\begin{equation}
\gamma_{\mathrm{CE}} \equiv \frac{\tau_{\mathrm spin-up}}{\tau_{\mathrm
decay}}\\
\end{equation}
where $\tau_{\mathrm{decay}}$ represents the decay timescale of the
planet's orbit, and $\tau_{\mathrm spin-up}$ is the spin-up timescale of
the envelope. The decay timescale is given by 
\begin{equation}
\tau_{\mathrm{decay}} = \Bigl|\frac{a}{\d a/\d t}\Bigr| \approx \frac{ \G
M_{\mathrm{bd}} M(a)}{2 a} \times \frac{1}{F_{\mathrm{drag}} V}\ ,
\end{equation}
where $M(a)$ the mass interior to $a$ and $F_{\mathrm{drag}}$ is the
gravitational drag force that causes the spiralling-in. Its expression is
approximately given by
\begin{equation}
F_{\mathrm{drag}}  \approx \xi \pi R_a^2 \rho V^2\ ,
\end{equation}
where $\xi$ is a factor of order $2-5$ for supersonic flows (e.g. Shima et
al. 1985), $R_a \approx 2\G 
M/V^2$ is the accretion radius, $V$ the relative velocity between the
planet and the envelope and $\rho$ the local density. Finally, the spin-up
timescale $\tau_{\mathrm{spin-up}}$ can be approximated by
\begin{equation}
\tau_{\mathrm{spin-up}} = \frac{I(a)V}{a^2 F_{\mathrm{drag}}}\ ,
\end{equation}
where $I(a)$ is the moment of inertia of the envelope interior to radius
$a$. In the computations for our presented models, we assumed that the
relative velocity $V$ is Keplerian and impose $\xi = 4$ as in Livio \&
Soker (1988). 
\begin{figure}
\psfig{file=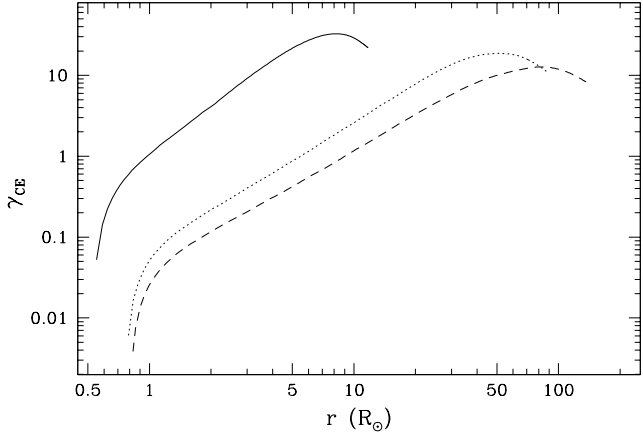,width=\columnwidth} 
\caption[]{The parameters $\gamma_{\mathrm{CE}}$ as a function of the radius
for the three initial models (see text). In the expression for
$\tau_{\mathrm{decay}}$, we took $M_{\mathrm{bd}}= 10
M_{\mathrm{Jup}}$. The solid, dotted and dashed lines correspond to the
initial models \#A, \#B and \#C, respectively.}
\label{beta}
\end{figure}
Figure \ref{beta} depicts the profiles of $\gamma_{\mathrm{CE}}$ for the 3
initial models.  Although these profiles are crude estimates, they suggest
that a strong differential rotation is likely to take place at the base of
the convective envelope, in the regions where $\gamma_{\mathrm{CE}}<1$. The
shear can then sustain a dynamo action and lead to a high X-ray
emission (e.g. Regos \& Tout 1995).
Furthermore, the profiles of $\gamma_{\mathrm{CE}}$ indicate that
for less evolved configurations the outer layers are not spun-up, and
therefore that it is possible to obtain X-ray emission from apparently
slowly rotating stars as observed among $\beta$ Ceti-like stars. 
It is difficult to estimate for how long the slow rotation will be
maintained at the surface of the star since such a study would require to
take into account the angular momentum transport in the convective envelope
coupled with the presence of a strong magnetic field.  Additionally the
outcome depends on the evolutionary status of the star. We expect however
that the spin up the envelope will take longer in younger and less
centrally condensed stars for which $\gamma_{\mathrm{CE}}>1$ in a larger
fraction of the envelope. In some configurations magnetic braking may not
allow a significant spin-up of the outer layers. Some
indications that differential rotation may persist in the interior of stars
are suggested by observations. Indeed, in the HR diagram, the rotational
($\S$\ref{rotation}) and X-ray dividing lines are shifted with respect to
one another. Thus, stars located between these two lines present both a
magnetic activity and a slow rotation rate. If this shift between the
dividing lines corresponds effectively to an evolutionary sequence, this
indicates either that the rotational braking timescale is shorter than the
decay timescale associated with the dynamo activity or, alternatively, that
differential rotation is still maintained in the stellar interior and
sustains a dynamo activity.

To summarize this section, the deposition of angular momentum by a
spiralling-in planet/brown dwarf can produce a strong differential rotation
at the base of the convective envelope. We then suspect that in this region
a magnetic field can be generated via an $\alpha-\Omega$ dynamo and that
this may ultimately result in an enhanced X-ray activity. Note that the
magnetic field can also be used to brake the star. Finally, we would like
to mention that the shear may trigger some instabilities that could give
rise to extra mixing processes (see $\S$\ref{lirich}).

\subsection{The ejection of shells}
\label{shell}

The increase in the mass loss rate during the accretion phase can lead to
the formation of a shell around the star, and that shell can be expected to
emit in the infrared (IR). More quantitatively, the mass loss rate can be
increased by a factor of 2 to $\sim 100$ and a mass of the order
$10^{-6}\Msun \la M_{\mathrm eject} \la 10^{-3}\Msun$ can be ejected. These
numbers depend however on the evolutionary status of the star, the duration
of the accretion event and the accretion rate.
\begin{figure}
\psfig{file=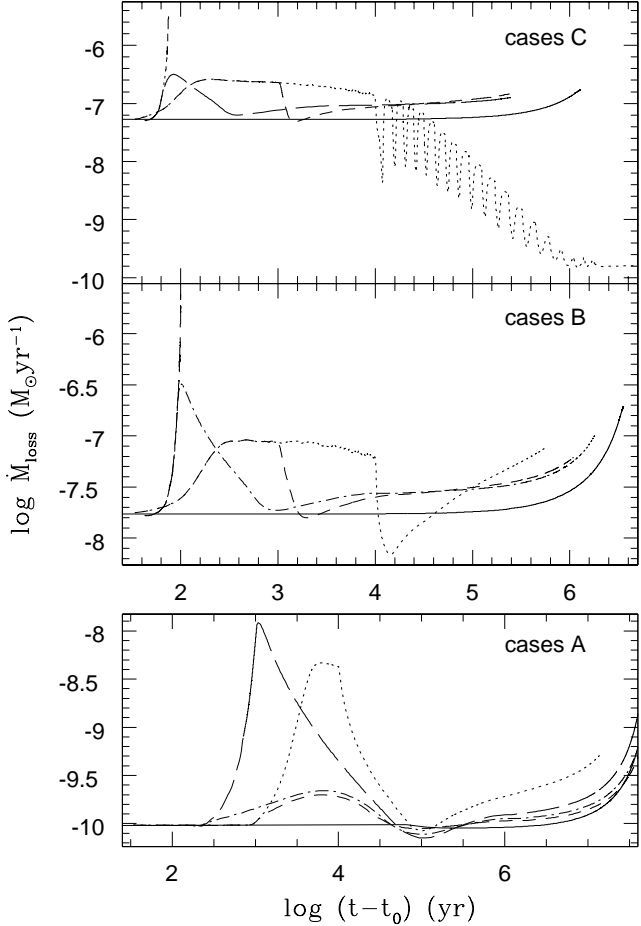,width=\columnwidth} 
\caption[]{The evolution of the mass loss rate is shown for our different
models. The solid lines refer to the standard evolution. In the lower
panel, the dotted, short-dashed, long-dashed and dotted-dashed lines
correspond to cases A0, A1, A2 and A3, respectively. The same lines refer
to cases B0, B1, B2 and B3 in the middle panel and to cases C0, C1, C2 and
C4 in the upper panel, respectively. The peaks in \mloss in the upper panel
are due to thermal pulses.}
\label{mloss}
\end{figure}
Figure \ref{mloss} demonstrates the very strong dependence of the mass loss
rate on the accretion rate and on the initial model. The higher the
accretion rate, the larger is the increase in the mass loss
rate. Similarly, the more evolved the star, i.e. the lower its surface
gravity, the easier it is to remove material from its surface. Figure
\ref{mloss} also indicates that the rate of increase in \mloss\ is faster
for older stars. This is an illustration of the thermal response of the
star since \mloss\ is inversely proportional to the Kelvin Helmholtz
timescale ($\Mloss \propto M/\tau_{\mathrm KH}$). Finally, we would like to
emphasize that in the high accretion rate cases (cases B2 and C2), even a
relatively small planet of a few Jupiter masses can increase the mass loss
rate by about 2 orders of magnitude in a few hundred years!

It is worth noting that in the estimates of the mass loss rate, the effects
of rotation have been neglected. The deposition of the orbital angular
momentum into the envelope can lower the effective surface gravity and
facilitate mass ejection; it can also influence of geometry of the
ejection. In particular, if the rotational velocity of the star exceeds
some fraction of the break-up speed, then the ejected material could form
an equatorially compressed outflow (e.g. Bjorkman \& Cassinelli 1993,
Owocki et al. 1994, Livio 1997). The peculiar circumstellar environment of
U Equulei could be consistent with such a scenario (Barnbaum et al. 1996).

The IR emission associated with giant stars is not easy to explain
(Zuckerman et al. 1995), while in main sequence stars (Vega-like stars) the
emission most likely arises from the dust associated with remnants of the
star formation process or with small bodies like comets or planetesimals
(e.g. Aumann et al. 1984, Telesco \& Knacke 1991, Backman \& Paresce
1993). In an extensive survey of luminosity class III stars, Plets et
al. (1997) report that $\sim 8$\% of the giants in their sample have IR
excesses. They conclude that the IR emission is unlikely to be due to a
present-day mass loss episode because, as they argue, $(i)$ in a standard
evolution, mass loss is predominantly efficient during a He shell flash so one
would expect the stars to be more clustered at higher luminosities in the
HR diagram and $(ii)$ the dust emission is too cold for a recent
ejection. They thus attribute the IR excess to a reheated dusty disk, as in
Vega-like stars. Instead, we propose that the ejection of a shell resulting
from an accretion event can account for the IR emission. Indeed, this
scenario does not imply a clustering of stars at the tip of the RGB since
the accretion event can, in principle, take place any time once the star
leaves the main sequence. Early engulfing could also account for the cold
dust emission. Moreover, the IR excess is often too large to be attributed
to a Vega-like phenomenon, since during the main sequence phase, most of
the circumstellar debris will probably be evaporated (Zuckerman et
al. 1995). An important test of our scenario can be provided by measuring
the \chem{7}Li abundance for example, since this element should, at some
level, accompany the accretion process (see $\S$\ref{lirich}).  \\ The
detection of a detached shell around the J-type carbon star Y Canum
Venaticorum (Izumiura et al. 1996) is also interesting since this star is
probably in an early evolutionary stage, on the RGB or on the He
core-burning stage. The mass loss rate at the time of shell formation and
the mass of the shell are estimated to be in the range $(7-20)10^{-6}$\myr
and $(4-10)10^{-2}$\msun, respectively. The present mass loss rate
estimated from CO observations (e.g. Olofsson et al. 1993) is about
$10^{-7}$\myr, two orders of magnitude smaller than the past value!
Therefore, if the RGB status of Y CVn is confirmed, the existence of a
massive shell around this star may be a signature of an accretion event,
since no other straightforward model can account for these observations. In
this context, the massive shell and high mass loss rate at the time of
ejection would suggest a very high accretion rate ($\Macc \ga
10^{-4}$\myr).

Shells have also been observed around many AGB stars (e.g. Olofsson et
al. 1992, 1996, Bujarrabal \& Cernicharo 1994), but such structures can
result from episodic high mass-loss events triggered by helium shell
flashes (e.g. Olofsson et al. 1990, Izumiura et al. 1997). It is therefore
difficult to identify the signature of an accretion event, although the
presence of a companion is frequently invoked for the shaping of planetary
nebulae (e.g. Soker 1997, Livio 1997). Finally, we should mention that in
model C0 the star experiences a series of thermal pulses before reaching
the horizontal branch. From, the simulations we can estimate that during
every interpulse, a mass of the order of $4\times 10^{-4}$\msun\ is ejected
from the surface of the giant. These episodic ejections would generate
distinct shells, separated from each other by a few thousand
years. However, the probability that a massive planet would be engulfed at
the short-lived tip of the giant branch is rather small.

\subsection{Influence on the morphology of globular clusters}
\label{morpho}

The distributions of stars along the horizontal branch (HB) of globular
clusters varies from one cluster to another. In some clusters, the
population of HB stars exhibits an extended blue tail while others contain
only a red HB. It was first recognized by Sandage and Whitney (1967) that
the metallicity is the first parameter influencing the morphology of the
HB. Generally, more metal-rich clusters have red HBs, while metal-poor
clusters have bluer HBs. But, for globular clusters of the same
metallicity, the HB morphology can vary widely. Thus some ``second
parameter'' other than metallicity must be affecting the HB
morphology. Among potential second parameters, it has been shown that age
(e.g. Lee et al. 1994), mass loss (e.g. D'Cruz et al. 1996), He mixing
(Sweigart 1997), CNO enhancement (Dorman et al. 1991), core rotation, or
structural properties of the globular clusters such as concentration and
density can also affect the HB morphology. Nevertheless the discovery of
gaps in the blue tail of the HB (e.g. Walker 1992, Catelan et al. 1997,
Ferraro et al. 1997) is still considered puzzling.

Recently, Soker (1998b) suggested that planetary systems can influence the
HB morphologies. He proposed that the interaction of a RGB star with a
giant planet/brown dwarf can affect the mass loss rate and thus the
subsequent location of the star on the HB. By depositing angular momentum
and energy into the envelope, additional mass can be removed from the star
during the RGB. The ``peeled off'' star will then have a higher effective
temperature and will appear bluer in the HR diagram (e.g. Castellani \&
Castellani 1993, Dorman et al. 1993). In particular, Soker (1998b) argued
that, depending whether the planet survives the spiralling-in, collides with
the stellar core or evaporates in the envelope, one can account for the 3
gaps observed by Sosin et al. (1997) in the blue tail of NGC 2808.

Our study introduces a quantitative element into the scenario proposed by
Soker. Our computations indeed indicate that a significant increase in the
mass loss rate accompanies the accretion process, especially if the star is
evolved and the accretion rate is high.  The simulations also show that,
when the star has relaxed, the mass loss rate remains higher than in a
standard evolution. We estimated that when a massive planet is accreted
(not a brown dwarf), the mass loss rate is enhanced by a factor of $1.3-2$
after the termination of the accretion process. This is equivalent to
setting Reimer's parameter $\eta_R$ to a value close to $0.7-1.0$ (instead
of 0.5) in a standard evolution. If we now refer to the computations of
Castellani \& Castellani (1993) for such values for $\eta_R$, we conclude
that stars accreting a planet at the bottom of the RGB, will actually be
extremely blue on the horizontal branch. On the other hand, if the
accretion occurs later, the effects are difficult to quantify because of
the strong sensitivity of our computations to the accretion rate and the
planet mass. The effects can be moderate, as in cases C2 or B2, or
dramatic, as in cases B3 or C4. The possibility that a planet accretion
would modify the evolutionary age of the star and thus affect the
morphology of the HB is improbable.  Indeed our initial models are
relatively old ($t_0 > 12$\,Myr) and the duration of the red giant phase is
relatively short compared to its age. Also, the accretion of a planet does
not modify substantially the mass of the star therefore we don't expect the
stellar evolutionary timescales to be significantly modified.

Sweigart (1997) suggested that an extra mixing driven by internal rotation
could transport helium from the HBS into the convective envelope. This
would increase the RGB tip luminosity and hence the mass loss, leading to
the formation of blue HB stars. Sweigart and Catelan (1998) also showed
that {\em only} non-canonical mechanisms invoking deep helium mixing,
internal rotation or a high initial helium abundance can reproduce the
distribution of the blue HB stars in NGC 6388 and NGC 6441. However, these
authors did not specify the source of the angular momentum required for the
rotation scenario nor how the deep He-mixing process is activated. The
accretion scenario can provide some support to these non-canonical
processes, since it both offers a natural source of angular momentum for
the core and it can be responsible for the generation of strong
instabilities as the planet collides with the core ($\S$\ref{xray}). Note
also that some rapid rotators are observed in the blue side of the HB
(e.g. Harris \& McClure 1985).

Concerning the gaps, recent observations (Ferraro et al. 1998) indicate
that they are not randomly distributed along the HB but rather located at
similar positions in the color-magnitude diagram. Aside from the
star-planet interaction hypothesis (Soker 1998b), it has been suggested
that some poorly specified physical processes must operate during the RGB
to increase the mass loss rate (Ferraro et al. 1998) or that gaps are a
consequence of a bimodal or multimodal mass distribution of the globular
clusters (e.g. Catelan et al. 1998). Let us therefore return to the planet
accretion scenario and analyze the implications of this assumption. We have
seen that depending on the evolutionary status of the star at the time of
engulfing (i.e. on the initial planet separation), different amounts of
mass can be lost and thus different positions in the HB can be
reached. Assuming for the moment that planets form preferentially around
solar-type stars (as it has been observed so far), the similarity in the
positions of the gaps would suggest that planets orbit at preferred
distances to the star. A theory suggesting that this is indeed the case
exists (Notalle 1996, Notalle et al. 1997 and references therein).  If this
is true, planets would be swallowed only when the solar-type star reaches a
certain radius (or equivalently an evolutionary status), determined
approximatively by the orbital separation of the planet. In this
hypothesis, the accretion process would modify selectively the mass loss
rate and thus the location of the star on the HB.\\ However, the question
of why the star-planet interaction process acts only sporadically and does
not appear as a general phenomenon affecting the morphology of all the
clusters would remain puzzling. To attempt to answer this question, one
needs a better understanding of the processes involved in planetary
formation. For example, metallicity and other globular cluster properties
such as density may be important factors that can influence the formation
of planets and they are still poorly understood.  To conclude this section,
the accretion of planets by giant stars can potentially influence the mass
loss rate and thus modify the position of the star on the HB. However, it
is not clear at present that the gaps observed in the blue tail of some
globular clusters are really the signature of such
accretion events.

\subsection{The lithium rich G and K giants}
\label{lirich}

The detection of lithium in the spectra of G and K giants (e.g. Wallerstein
\& Sneden 1982, Brown et al. 1989, Gratton \& D'Antona 1989, Pilachowski et
al. 1990, Castilho et al. 1995, de la Reza \& da Silva 1995, Castilho et
al. 1998) poses a problem concerning the origin of this element in these
stars since, in standard evolution, lithium is entirely depleted during the
first dredge up. Several mechanisms have been suggested to explain the Li
enrichment and we refer to Gratton and D'Antona (1989) for a more complete
discussion. Briefly, it has been proposed for example that Li could be
preserved in the atmosphere of giant stars, but the low
\chem{12}C/\chem{13}C ratio observed in these stars (e.g. Pilachowski et
al. 1990, de la Reza et al. 1995) indicates that deep mixing (and thus Li
burning) has occurred. This represents therefore a failure of this
scenario. Gratton and D'Antona (1989) considered the possibility that the
giant's envelope could be contaminated by a nova exploding companion.
However, the non detection of a white dwarf in the vicinity of Li-rich
giants (e.g. de la Reza \& da Silva 1995) weakens this hypothesis.  The
possibility that Li could be produced by the Cameron-Fowler mechanism
(1971) has also been investigated. This mechanism, which consists of the
rapid dredge up of \chem{7}Be and Li production through the
\chem{7}Be($e^-, \nu$)\chem{7}Li reaction, is known to operate in AGB stars
of 4-6\msun\ for which the convective timescale is short compared to the
\chem{7}Be destruction timescale (Sackmann \& Boothroyd
1992). Unfortunately, this process is inefficient in low luminosity stars,
unless one invokes some extra mixing processes (we will return to this
point later). Finally, and we will develop this point below, several
authors (e.g. Alexander 1967, Brown et al. 1989, Gratton and D'Antona 1989)
proposed that the engulfing of a planet or a brown dwarf could be
responsible for the high Li abundance.

\begin{figure}
\psfig{file=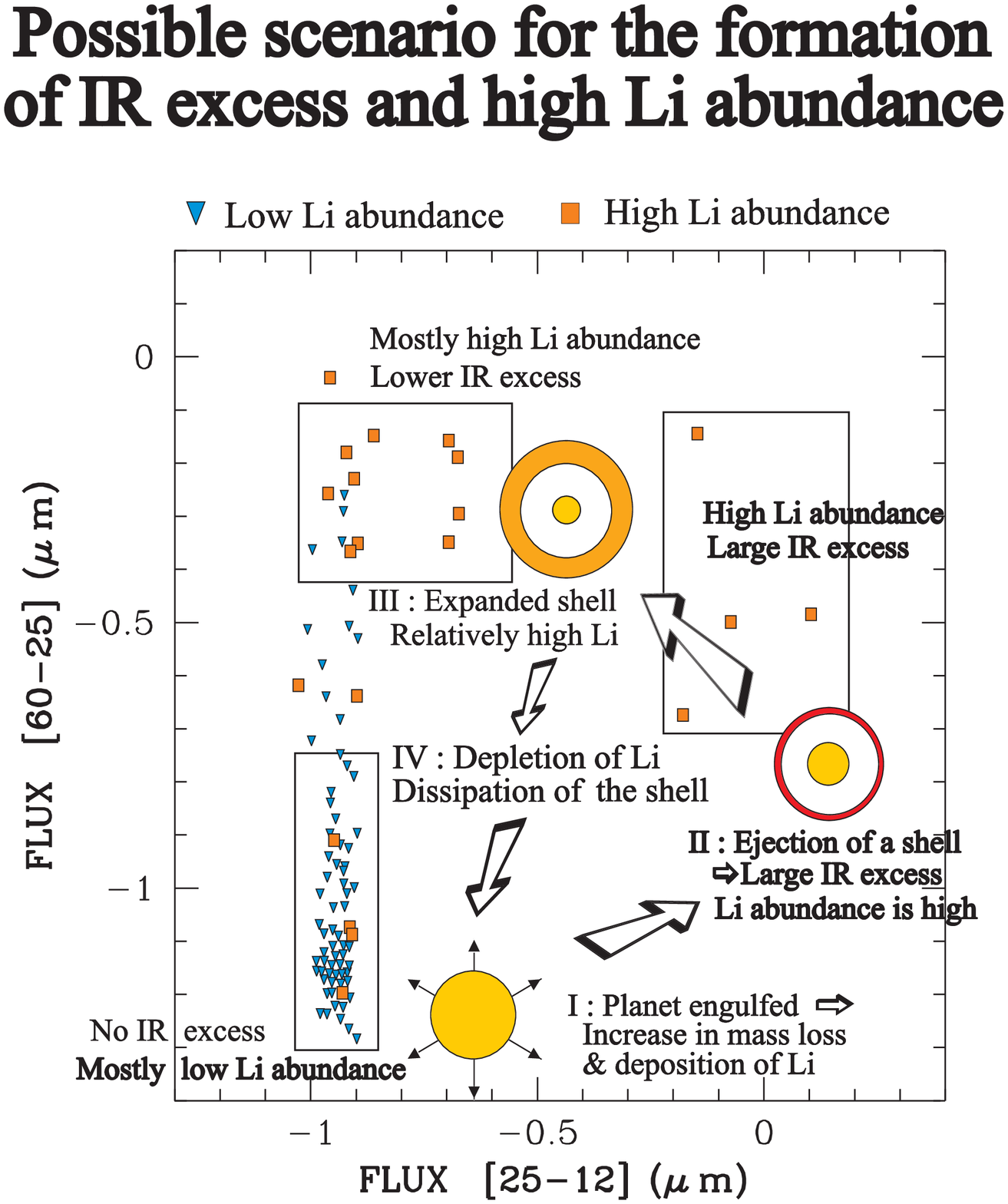,width=\columnwidth} 
\caption[]{The schematic evolution in an IRAS color-color diagram of a K
giant star engulfing a planet or a brown dwarf. (By definition,
[$\lambda_1-\lambda_2]=\log \lambda_2S_{\lambda_1} - \log \lambda_1
S_{\lambda_2}$, where $S_\lambda$ is the IRAS density flux.) This figure
was adapted from de la Reza et al. (1996) and is explained in the text.}
\label{dela}
\end{figure}

Based on IRAS observations of G and K giants, de la Reza et al. (1996,
1997) concluded that {\it almost all} the Li rich G and K giants have an IR
excess compatible with the presence of a circumstellar shell. To account
for these observations, they constructed a model in which Li is produced
inside the star by an extra mixing process, the ``cool bottom burning''
(Wasserburg et al. 1995). This process allows the fast transport of
\chem{7}Be from the HBS into the convective envelope where \chem{7}Be then
decays to produce \chem{7}Li. Furthermore, they assumed that the mechanism
for Li enrichment is accompanied by a sudden increase in the mass loss rate
and thereby the ejection of a shell. Following the evolution of a thin
expanding circumstellar shell, they could reasonably account for the IR
emission and Li abundance distribution of G and K giants in a color-color
diagram. However, de la Reza et al. remained relatively vague concerning
the mechanism that is responsible for mass loss; they assumed that it is a
consequence of the Li enrichment process. Instead, we propose here that the
accretion of a planet or a brown dwarf can naturally and consistently
explain all of the observed features. We describe our scenario in the next
paragraph and illustrate it in Fig. \ref{dela} which is an adaptation of
Fig. 2 of de la Reza et al. (1996).

Figure \ref{dela} represents a color-color diagram based on the IRAS
density fluxes at 12, 25 and 60 $\mu$m. In this diagram, de la Reza et
al. (1996, hereafter DDD) identified 3 regions, populated respectively by
normal K giants with no IR excess (region $I$), Li rich giants with a large
IR excess (region $II$), and relatively Li rich giants with a moderate IR
excess (region $III$).  When the accretion process starts, Li is
progressively deposited in the stellar envelope and the mass loss rate
increases. Rapidly, a shell is ejected and the giant moves quickly from
region I to region II. In their shell model, DDD found that several hundred
years are required to reach region $II$. This timescale gives an estimate
of the duration of the accretion process. Then, Li starts to be depleted in
the envelope while the shell expands and cools down. The 25$\mu$m flux
decreases and the star now settles in region $III$. DDD estimated that it
takes approximatively $10^4$yr to move from region $II$ to $III$. Finally,
in about $8\,10^4$yr, the circumstellar shell dissipates, Li is entirely
depleted and the star returns to region $I$ as a normal K giant. Assuming a
wind velocity of 2 \kms and a mass loss rate of $5\,10^{-8}$\myr, DDD found
that $\sim 10^{-5}$\msun\ was ejected in the shell, in agreement with our
models. In a second publication, de la Reza et al. (1997) completed their
study by presenting new observations. They showed that Li-rich and Li-poor
stars are mixed in the color-color diagram.  They interpreted these
observations as being due to differences in the Li destruction
timescales. In our model, this is explained naturally if the amount of Li
deposited into the star, or equivalently the mass of the planet, is
different from one star to another.

The mechanism that leads to the Li enrichment must also account for its
rapid depletion, which takes place in $\sim 10^5$yr. In our computations
without diffusion, Li is maintained at the surface of the star but
Charbonnel (1995) showed that this element can be rapidly depleted in red
giants due to non-standard mixing processes. In particular, the model
developed by Zahn (1992) showed that the efficiency of the mixing is mainly
determined by the loss of angular momentum via a stellar wind and, even in
the absence of a high mass loss rate, the process takes place where there
is a strong gradient in $\omega$. He also pointed out that additional
mixing is expected in shell burning regions. Since all of these conditions
are satisfied in the context of the accretion process (increase in the mass
loss rate, deposition of angular momentum and a strong gradient of $\omega$
in the dissipation region as described in $\S$\ref{xray}) and as a part of
the fact that the star is on the RGB (the presence of the HBS), we have
every reason to believe that the mechanism proposed by Zahn (1992), and
that successfully explains some peculiar anomalies in red giant stars
(Charbonnel 1995), could be responsible for the fast Li depletion following
a short accretion phase. Moreover, observations of the
\chem{12}C/\chem{13}C ratio in some Li rich giants (da Silva \& de la Reza
1995) indicate that some extra mixing is operating since the carbon isotope
ratio in these stars is lower than the standard value. \\ Our scenario does 
encounter some difficulties in explaining the very high Li abundances
[log\,$\epsilon$(Li)$\ga 2.8$] found in some giants (e.g. Gratton \&
D'Antona 1989, de la Reza \& da Silva 1995, Fekel et al. 1996). 
This would require the deposition of a huge amount of lithium that can
only be achieved either by the accretion of numerous planets such that 
$M_{\mathrm acc} \approx M_{\mathrm env}$ or if the Li
abundance in the brown dwarf is extremely high ($X^{\mathrm acc} \ga 10
\times X^{\mathrm env}$ for $M_{\mathrm acc} \simeq 0.1$\msun). The
accretion scenario does not provide a straightforward explanation but we 
can imagine that when the planet reaches the central region, the
dissipation process may generate some instabilities (due to the shear) that
could, in turn, trigger the ``cool bottom burning'' invoked by de la Reza et
al. (1996, 1997). The accretion process would then lead to some Li
production (as in more massive AGB stars).

De Medeiros et al. (1996a) reported that the majority of Li K giants are
normal rotators ($v\sin i \simeq 2-3$ \kms) although some exceptions exist
[e.g. HD 233517 (Fekel et al. 1996), HD 9746, HD 25893, HD 31993, HD 33798,
HD 203251 (Fekel \& Balachandran 1993), 1E 1751+7046 (Ambruster et
al. 1997)].  However, as pointed out in $\S$\ref{rotation}, high angular
velocities ($\sim 10$\% percent of the Keplerian velocity) can only be
achieved if a sufficiently massive planet ($M_{\mathrm{bd}} > 5
M_{\mathrm{Jup}}$) is accreted. For a typical 1.0\msun\ RGB star, with a
radius $R \sim 20$\rsun, the deposition of the orbital angular momentum of
a 5 $M_{\mathrm Jup}$ planet will ``only'' increase the star's rotational
velocity to $\sim 5.5$ \kms\ [Eq. (\ref{omega})].  Consequently, as long as
the mass of the engulfed body is low (a planet rather than a brown dwarf),
the observed angular momentum of Li rich giants is compatible with our
scenario. We should also note that due to the increased mass loss, magnetic
braking could also be operating (e.g. Leonard \& Livio 1995).  Finally, let
us point out that {\em all the Li rich giants are single stars}, hence
according to the prevailing common wisdom, they could possess planetary
systems.

To summarize this section, we propose that the accretion of a planet or a
brown dwarf is responsible for both the IR excess and the high Li abundance
present in almost all single Li rich G and K giants.

\section{Discussion and conclusion}
\label{conclusion}

In this paper, we investigated the effects of the accretion of a planet or
a brown dwarf by a solar type star located on the red giant branch.  Our
computations show that the deposition of mass (potential energy) produces a
substantial expansion of the envelope and, for higher accretion rates,
nuclear burning at the base of the convective envelope can take place. The
simulations show that the results are a sensitive function of the accretion
rate and the structure of the star. Generally, the higher the accretion
rate and/or the more evolved the star, the larger is the expansion. After
the end of the accretion process, the star relaxes and resumes a standard
evolution.  However, in one simulation (case C0), the star avoided the He
flash but instead developed thermal instabilities. For this model, after a
phase of thermal pulses, He burning was finally ignited at the center of
the star.

We discussed the possible observational signatures of the accretion of a
planet/brown dwarf. In particular (1) we showed that the rotational
velocity of the star can be substantially increased, provided that the
accreted planet/brown dwarf is sufficiently massive. In this respect the
accretion of a brown dwarf could account for the fast rotation of FK Comae
stars and more generally, the deposition of angular momentum via this
process can explain unusually fast rotation rates. (2) Investigating the
X-ray properties of giant stars, we showed that a strong differential
rotation may take place at the bottom of the convective envelope. The
presence of shear in this region can turn-on the dynamo activity and this
may ultimately generate X-ray emission. We showed that under some
conditions the surface layers of the star could rotate slowly while a rapid
rotation is maintained in the core. (3) We also showed that the accretion
scenario can lead to the ejection of shells whose mass is in the range
$10^{-6}\Msun \la M_{\mathrm eject} \la 10^{-3}\Msun$. The IR excesses
observed in $\approx$8\% of luminosity class III stars could in part be due
to this process. (4) We argued that the increase in the mass loss rate can
also have indirect effects on the morphology of the horizontal branch and
in this respect, the accretion process is a potential second parameter
candidate. (5) We showed that the deposition of planet material in the
convective envelope can efficiently increase the \chem{7}Li surface
abundance and more generally enhance the stellar metallicity of giant
stars. (6) Finally, on a related topic, we proposed that the accretion of a
planet/brown dwarf can explain consistently both the IR excess and the
high Li abundance observed only in single red giant stars.

In terms of physical processes, we suggested that planet/brown dwarf
accretion may provide the triggering for a variety of processes. For
example, the disruption of the planet and the associated instabilities
could increase the He abundance in the giant's envelope and thus affect the
morphology of the HB as proposed by Sweigart (1997, $\S$\ref{morpho}). They
could also be responsible for the production of lithium through the
mechanism proposed by Wasserburg et al. (1995, $\S$\ref{lirich}).  Finally,
the rapid contraction of the star at the end of the accretion process could
generate some pulsations.

Observations indicate that several of potential different signatures of an
accretion event, e.g. a fast rotation rate, a significant X-ray emission, a
high Li abundance or a large IR excess, do not always occur
simultaneously. As we have shown, this is not surprising. First, depending
on the mass of the planet, the outcomes could be different. For example, a
small planet that evaporates in the envelope before reaching its base will
perturb the structure only slightly. In such a case, no shell will be
ejected but the deposition of angular momentum and matter in the surface
layers could still produce a relatively rapid rotation and a high lithium
abundance, as observed in some chromospherically active giants (Fekel \&
Balachandran 1993). Secondly, the timescales associated with the various
phenomena (IR excess, X-ray activity, rotational braking, particle
transport) can be different, and thus, not all the signatures will be
observed at the same time. For example, if the spin-down timescale is
longer than the timescale associated with the dissipation of the ejected
shell, one can account for the absence of IR excess in some fast
rotators. Thirdly, the consequences of the accretion process depend on the
evolutionary status of the star. Giant stars located at the bottom of the
RGB have much deeper convective envelopes than near the tip of the giant
branch. Thus, if we consider for example the deposition of Li in the
envelope, it is more difficult to obtain high Li abundances in younger red
giants (it is also be more difficult to eject a shell). Finally, our
simulations show that the results are sensitive to the physics of the
dissipation process, and more specifically, to the accretion rate and the
location of the accretion process. At present, both these physical
parameters are rather uncertain, and a better understanding of the
dissipation process is definitively required.

As a last point, we can attempt to use the expected observational
signatures to infer some estimate for the frequency of planets around
solar-type stars.  In a search for lithium in 644 G and K giants, Brown et
al. (1989) found that 2\% in the stars of their sample have
log\,$\epsilon$(Li)$\ge 1.5$ and 4\% have log\,$\epsilon$(Li)$> 1.3$. If we
adopt for our definition of Li rich giants the value log\,$\epsilon$(Li)$>
1.2$, we find that $\sim 8$\% of the stars surveyed by Brown et al.  belong
to this group. In a more recent survey, Castilho et al. (1998) report the
detection of 5 Li rich and 6 moderately Li rich giants among 164 stars,
indicating a proportion of Li rich giants of $\sim 3-7$\%. The study of
Plets et al. (1997) indicates that 8\% of luminosity class III giants have
unexpected IR excess. Finally, based on rotational velocity measurements of
$\sim 900$ G and K giants, De medeiros et al. (1996a) found that $\la 5\%$
of late G and K giants located beyond the rotational dividing line are fast
rotators with $v\,\sin\,i > 5$ \kms. Thus, if the IR excess and/or high Li
abundance and/or unusually high rotational velocity among G and K giants
result from the accretion of a planet/brown dwarf, the above numbers
suggest that at least 4-8\% of the single, solar-type stars must have a low
mass companion, either a brown dwarf or a planet. If we were to take
seriously the frequency of planets inferred by Soker (1997) to account for
the shaping of planetary nebulae ($\sim 55\%$ of all progenitors of masses
$\la 5$\msun), our deduced frequency would indicate that either a large
fraction of the planets do not interact during the red giant phase or, more
probably, that the observational signatures of the accretion event are very
short-lived compared to stellar evolutionary timescales.  Indeed, because
of the large radii reached by solar type stars during the RGB, some planets
will most likely be engulfed during the early giant phase rather than on
the AGB (although Jupiter will not be engulfed by our own sun). On the
other hand, if we compare the timescales associated with both the Li and IR
excesses ($\ga 10^5$yr) to the red giant lifetime (between $6\,10^6$ and
$5\,10^8$yr for giants with masses between 1.0 and 2.5\msun), we conclude
that the probability of detecting an accretion event is such that we are
likely to miss most candidates.

\section*{Acknowledgments}
 
LS acknowledges support from the Director's Discretionary Research Fund at
STScI and thanks the STScI for its hospitality. This work has been
supported in part by NASA grants NAGW-2678, G005.52200 and G005.44000.  The
computations presented in this paper were performed at the ``Centre
Intensif de Calcul de l'Observatoire de Grenoble''.


\begin{thebibliography}{99}
  
\bibitem{} 
Alexander J.B., 1967, Observatory, 87, 238 

\bibitem{} 
Ambruster C.W., Fekel F.C., Guinan E.F., Hrivnak B.J., 1997,ApJ, 479, 960 
 
\bibitem{} 
Anders E., Grevesse N., 1989, Geochim. Cosmochim. Acta, 53, 197

\bibitem{} 
Aumann H.H., Gillett F.C., Beichmann C.A. et al., 1984, ApJ,
278, L23 

\bibitem{} 
Ayres T.R., Simon T., Stern R.A., Drake S.A., Wood B., Brown A., 1998, ApJ,
496, 428 

\bibitem{} 
Backman D.E., Paresce F., 1993, in Protostars and Planets III,
eds. E.H. Levy, J.I. Lunine, The University of Arizona Press, p. 1253 

\bibitem{} 
Barnbaum C., Morris M., Kahane C., 1995, ApJ, 450, 862
 
\bibitem{} 
Basri G., Marcy G.W., Graham J.R., 1996, ApJ, 458, 600

\bibitem{} 
Bjorkman J.E., Cassinelli J.P., 1993, ApJ, 409, 429

\bibitem{} 
Bopp B.W., Stencel R.E., 1981, ApJ, L31
 
\bibitem{} 
Brown J.A., Sneden C., Lambert D.L., Dutchover E.Jr., 1989, ApJS, 71, 293

\bibitem{} 
Bujarrabal V., Cernicharo J., 1994, A\&A, 288, 551
 
\bibitem{} 
Butler R.P., Marcy G.W., 1996, ApJ, 464, L153

\bibitem{} 
Butler R.P., Marcy G.W., Vogt S.S., Apps K., 1998, PASP, 110, 1389
 
\bibitem{} 
Cameron A.G.W., Fowler W.A., 1971, ApJ, 164, 11
 
\bibitem{} 
Castellani M., Castellani V., 1993, ApJ, 407, 649
 
\bibitem{} 
Castilho B.V., Barbuy B., Gregorio-Hetem J., 1995, A\&A, 297, 503
 
\bibitem{} 
Castilho B.V., Gregorio-Hetem J., Spite F., Spite M., Barbuy B., 1998, A\&AS,
127, 139 

\bibitem{} 
Catelan M., Borissova J., Sweigart A.V., Spassova N., 1998, ApJ, 494, 265

\bibitem{} 
Charbonnel C., 1995, ApJ, 453, L41
 
\bibitem{} 
Cochran W.D., Hatzes A., Marcy G.W., Butler R.P., 1997, ApJ, 483, 457

\bibitem{} 
Cohen J.G., McCarthy J.K., 1997, AJ, 113, 1353

\bibitem{} 
D'Cruz N.L., Dorman B., Rood R.T., O'Connell R.W., 1996, ApJ, 466, 359

\bibitem{} 
Delfosse X., Forveille T., Mayor M., Perrier C., Naef D., Queloz D., 1998,
A\&A, 331, 581

\bibitem{} 
de la Reza R., da Silva L., 1995, ApJ, 439, 917
 
\bibitem{} 
de la Reza R., da Silva L., Barbuy B., 1995, ApJ, 448, L41
 
\bibitem{} 
de la Reza R., Drake N.A., da Silva L., 1996, ApJ, 456, L115, DDD
 
\bibitem{} 
de la Reza R., Drake N.A., da Silva L., Torres C.A.O., Martin E.L., 1997,
ApJ, 482, L77
 
\bibitem{} 
de Medeiros J.R., Melo C.H.F., Mayor M., 1996a, A\&A, 309, 465
 
\bibitem{} 
de Medeiros J.R., da Rocha C., Mayor M., 1996b, A\&A, 314, 499
 
\bibitem{} 
Dorman B., Lee Y.-W., VandenBerg D.A., 1991, ApJ, 366, 115
 
\bibitem{} 
Dorman B., Rood R.T., O'Connell R.W., 1993, ApJ, 419, 596
  
\bibitem{} 
Fekel F.C., 1988, in Adecade of UV Astronomy with {\sl IUE} Satellite,
ed. E.J. Rolfe (ESA Publications, Noordwijk, The Netherlands), Vol. 1,
p. 331 
  
\bibitem{}
Fekel F.C., Balachandran S., 1993, ApJ, 403, 708

\bibitem{} 
Fekel F.C., Marschall L.A., 1991, AJ, 102, 1439
 
\bibitem{} 
Fekel F.C., Webb R.A., White R.J., Zuckerman B., 1996, ApJ, 462, L95
 
\bibitem{} 
Ferraro F.R., Paltrinieri B., Fusi Peci F., Rood R.T., Dorman B., 1998,
ApJ, 500, 311
 
\bibitem{} 
Forestini M., Charbonnel C., 1997, A\&AS, 123, 241

\bibitem{} 
Gonzalez G., 1997, MNRAS, 285, 403
 
\bibitem{} 
Gonzalez G., 1998, A\&A, 334, 221
 
\bibitem{} 
Gratton R.G., D'Antona F., 1989, A\&A, 215, 66

\bibitem{} 
Gray D.F., 1981, ApJ, 251, 155

\bibitem{} 
Gray D.F., 1982, ApJ, 262, 682

\bibitem{} 
Gray D.F., 1986, Highlights Astr, 7, 411

\bibitem{} 
Gray D.F., 1989, ApJ, 347, 1021

\bibitem{} 
Gray D.F., 1991, in Angular Momentum Evolution of Young Stars, eds
S. Catalano, J.R. Stauffer, Kluwer, Dordrecht, p. 183

\bibitem{} 
Gray D.F., Nagar P., 1985, ApJ, 298, 756

\bibitem{} 
Haisch B., Schmitt J.H.M.M., Fabian A.C., 1992, Nature, 360, 239

\bibitem{} 
Haisch B., Schmitt J.H.M.M., Rosso A.C., 1991, ApJ, 383, L15

\bibitem{} 
Harpaz A., Soker N., 1994, MNRAS, 270, 734
 
\bibitem{} 
Harris H.C., McClure R.D., 1985, PASP, 97, 261

\bibitem{} 
H\"unsch M., Schr\"oder K.-P., 1996, 309, L51

\bibitem{} 
H\"unsch M., Schmitt J.H.M.M., Schr\"oder K.-P., Reimers D., 1996,
A\&A, 310, 801

\bibitem{} 
H\"unsch M., Schmitt J.H.M.M., Schr\"oder K.-P., Zickgraf F.-J., 1998,
A\&A, 330, 225

\bibitem{} 
Iben I.Jr., Livio M., 1993, PASP, 105, 1373

\bibitem{} 
Izumiura H., Hashimoto O., kawara K., Yamamura I., Waters L.B.F.M., 1996,
A\&A, 315, L221

\bibitem{} 
Izumiura H., Waters L.B.F.M., de Jong T., Loup C., Bontekoe Tj.R., Kester
D.J.M., 1997, A\&A, 323, 449

\bibitem{} 
Kraft R.P., 1970, in Spectroscopic Astrophysics, ed. G.H. Herbig (Berkeley,
University of California Press), p. 385 

\bibitem{} 
Laughlin G., Adams F.C., 1997, apJ, 491, L51

\bibitem{} 
Lee Y.-W., Demarque P., Zinn R., 1994, ApJ, 423, 248
 
\bibitem{} 
Leonard P.T.J., Livio M., 1995, ApJ, 447, L21
 
\bibitem{} 
Lin D.N., Bodenheimer P., Richardson D.C., 1996, Nature, 380, 606
 
\bibitem{} 
Livio M., 1997, Space Sci. Rev., 82, 389
 
\bibitem{} 
Livio M., Soker N., 1984, MNRAS, 208, 763
 
\bibitem{} 
Livio M., Soker N., 1988, ApJ, 329, 764

\bibitem{} 
Maggio A., Favata F., Peres G., Sciortino S., 1998, A\&A, 330, 139
 
\bibitem{} 
Marcy G.W., Butler R.P., Vogt S.S., Fischer D., Lissauer J.J., 1998, ApJ,
505, L147 

\bibitem{} 
Mayor M., Queloz D., 1995, Nature, 378, 355

\bibitem{} 
Notalle L., 1996, A\&A, 315, L9

\bibitem{} 
Notalle L., Schumacher G., Gay J., 1997, A\&A, 322, 1018

\bibitem{} 
Olofsson H., Bergman P., Eriksson K., Gustafsson B., 1996, A\&A, 311, 587

\bibitem{} 
Olofsson H., Carlstr\"om U., Eriksson K., Gustafsson B., Willson L.A.,
1990, A\&A, 230, L13 

\bibitem{} 
Olofsson H., Carlstr\"om U., Eriksson K., Gustafsson B., 1992, A\&A, 253,
L17 

\bibitem{} 
Olofsson H., Eriksson K., Gustafsson B., Carlstr\"om U., 1993, ApJS, 87,
267 

\bibitem{} 
Owocki S.P.,  Cranmer S.R., Blondin J.M., 1994, ApJ, 424, 887

\bibitem{} 
Peterson R.C., Rood R.T., Crocker D.A., 1995, ApJ, 453, 214

\bibitem{} 
Peterson R.C., Tarbell T.D., Carney B.W., 1983, ApJ, 265, 972

\bibitem{} 
Pilachowski C.A., Sneden C., Hudek D., 1990, AJ, 99, 1225

\bibitem{} 
Plets H., Waelkens C., Oudmaijer R.D., Waters L.B.F.M., 1997, A\&A, 323,
513  
 
\bibitem{} 
Rasio F.A., Tout C.A., Lubow S.H., Livio M., 1996, ApJ, 470, 1187
 
\bibitem{} 
Rebolo R., Martin E.L., Basri G., Marcy G., Zapatero-Osorio M.R., 1996,
ApJ, 469, L53
 
\bibitem{} 
Rebolo R., Zapatero-Osorio M.R., Martin E.L., 1995, Nature, 377, 129

\bibitem{} 
Regos E., Tout C.A., 1995, MNRAS, 273, 146

\bibitem{} 
Reimers D., 1975, Men. Soc. Roy. Sci. Li\`ege, 6th ser. 8, 369 

\bibitem{} 
Reimers D., H\"unsch M., Schmitt J.H.M.M., Toussaint F., 1996, A\&A, 310,
813 

\bibitem{} 
Rosner R., An C.-H., Musielak Z.E., Moore R.L., Suess S.T., 1991, ApJ, 327,
L91 

\bibitem{} 
Rosner R., Musielak Z.E., Cattaneo F., Moore R.L., Suess S.T., 1995, ApJ, 442,
L25

\bibitem{} 
Rucinski S.M., 1985, MNRAS, 215, 591

\bibitem{} 
Rucinski S.M., 1990, PASP, 102, 306

\bibitem{} 
Sackmann I.J., Boothroyd A.I., 1992, ApJ, 392, L71

\bibitem{} 
Sandage A.R., Whitney R., 1967, 150, 469

\bibitem{} 
Sandquist E., Taam R.E., Lin D.N.C., Burkert A., 1998, ApJ, 506, L65 

\bibitem{} 
Shima E., Matsuda T., Takeda H., Sawada K., 1985, MNRAS, 217, 367

\bibitem{} 
Siess L., Livio M., 1999, MNRAS, in press

\bibitem{} 
Siess L., Forestini M., Bertout C., 1997, A\&A, 326, 1001

\bibitem{} 
Simon T., Drake S.A., 1989, ApJ, 364, 303

\bibitem{} 
Soker N., 1997, ApJS, 112, 487

\bibitem{} 
Soker N., 1998a, ApJ, 498, 833

\bibitem{} 
Soker N., 1998b, in the Proceedings of the 10th Cambridge Workshop on Cool
Stars, Stellar Systems and the Sun, in press
 
\bibitem{} 
Soker N., Harpaz A., Livio M., 1984, MNRAS, 210, 189

\bibitem{} 
Sosin C., Dorman B., Djorgovski S.G., et al., 1997, 480, L35

\bibitem{} 
Stepien K., 1993, ApJ, 416, 368

\bibitem{}
Strassmeier K.G., Hall D.S., Barksdale W.S., Jusick A.T., Gregory W.H., 
1990, ApJ, 350, 367

\bibitem{}
Strassmeier K.G., Handler G., Paunzen E., Rauth M., 1994, A\&A, 281, 855 

\bibitem{}
Sweigart A.V., 1997, ApJ, 474, L23

\bibitem{}
Sweigart A.V., Catelan M., 1998, ApJ, 501, L63
 
\bibitem{}
Taam R.E., Bodenheimer P., 1989, ApJ, 337, 849
 
\bibitem{}
Telesco C.M., Knacke R.F., 1991, ApJ, 372, L29
 
\bibitem{}
Vitenbroek H., Dupree A.K., Gilliland R.L., 1998, AJ, 116, 2501
 
\bibitem{}
Wallerstein G., Sneden C., 1982, ApJ, 225, 577

\bibitem{} 
Walker, 1992, PASP, 104, 1063
 
\bibitem{}
Wasserburg G.J., Boothroyd A.I., Sackmann I.J., 1995, ApJ, 447, L37

\bibitem{} 
Webbink R.F., 1977, ApJ, 211, 486

\bibitem{} 
Welty A.D., Ramsey L.W., 1994, ApJ, 435, 848

\bibitem{} 
Zahn J.-P., 1992, A\&A, 265, 115

\bibitem{} 
Zuckerman B., Kim S.S., Liu T., 1995, ApJ, 446, L79

\end{thebibliography}
\end{document}